\documentclass[conf]{new-aiaa}
\usepackage{textcomp}
\usepackage[linesnumbered,ruled]{algorithm2e}
\SetKw{Break}{break}
\SetKw{Or}{or}

\newtheorem{remark}{Remark}

\usepackage{blindtext}
\usepackage{multicol}
\usepackage{verbatim}
\usepackage{xparse}

\usepackage{graphicx}

\usepackage[hyperref]{xcolor}

\usepackage{bm}

\usepackage{mathtools}
\usepackage{mleftright} 

\usepackage{stackengine}
\DeclareDocumentCommand \crule
{  m o }{
\stackengine{0.5pt}{\textcolor{black}{\rule{\IfValueTF{#2}{#2}{10pt}}{6pt}}}{\textcolor{#1}{\rule{\IfValueTF{#2}{#2-1pt}{9pt}}{5pt}}}{O}{c}{F}{F}{L}	
}

\usepackage{enumitem}
\usepackage{import}
\usepackage{amsmath}
\usepackage{siunitx}
\usepackage[c]{esvect}

\usepackage{interval} 

\usepackage{booktabs}
\usepackage{multirow}

\usepackage{cuted}
\usepackage{framed}
\usepackage[super]{nth}

\usepackage{cleveref}
\crefformat{equation}{(#2#1#3)}

\usepackage{subcaption}

\DeclareDocumentCommand \coordframe
{  m m m m }{
\left(#1\,;#2,#3,#4\right)
}

\DeclareDocumentCommand \transportmat
{  m m o }{
\mf{\IfNoValueTF{#3}{T}{#3 T}}[#1#2]
}

\DeclareDocumentCommand \statespace{m m m m o}{%
\ensuremath{{\def\arraystretch{1.3}\IfValueTF{#5}{#5}{}\left(\begin{array}{c|c}
    #1 & #2 \\ \hline
    #3 & #4
\end{array}\right)}}
}

\DeclareMathOperator{\card}{\textbf{card}}

\DeclareMathOperator*{\minimize}{\text{minimize}}

\DeclareDocumentCommand \rhinf
{  o o } {%
\mathbb{RH}_\infty\IfValueTF{#1}{^{#1 \times #2}}{}%
}

\newcommand{\ts}{%
T
}

\DeclareDocumentCommand \SmallBmatrix
{ m o } {
		\ensuremath{
		\begingroup
		\IfNoValueTF{\footnotesize}{#2}
		\arraycolsep=0.1\arraycolsep
		\renewcommand*{\arraystretch}{0.3}
		\begin{bmatrix}#1\end{bmatrix}
		\endgroup
	}
}

\newcommand\SmallMatrix[1]{{%
		\ensuremath{
			\begingroup
			\small
			\arraycolsep=0.1\arraycolsep
			\renewcommand*{\arraystretch}{0.3}
			\begin{matrix}#1\end{matrix}
			\endgroup
}}}

\DeclareDocumentCommand \iv
{  m o  } {%
{\IfNoValueTF{#2}{\footnotesize}{#2}
\arraycolsep=2pt\def\arraystretch{1}
\left[
\begin{matrix}
#1
\end{matrix}
\right]}%
}

\DeclareDocumentCommand \smax
{  m o o } {%
	\IfNoValueTF{#2}{%
	\overline{\sigma}\left(#1\right)%
	}{%
	\overline{\sigma}#2#1#3
	}
}
\DeclareDocumentCommand \smin
{  m o o } {%
	\IfNoValueTF{#2}{%
	\underline{\sigma}\left(#1\right)%
	}{%
	\underline{\sigma}#2#1#3
	}
}

\DeclareDocumentCommand \sf
{  m O{} o } {
	\boldsymbol{#1}_{#2}\IfNoValueTF{#3}{}{\{#3\}}%
}
\DeclareDocumentCommand \vf
{  m O{} o } {
	\mathbf{#1}_{#2}\IfNoValueTF{#3}{}{\{#3\}}%
}
\DeclareDocumentCommand \vfs
{  m O{} o } {
	\mathbf{#1}_{#2}\IfNoValueTF{#3}{}{\{#3\}}%
}
\DeclareDocumentCommand \mf
{  m O{} O{} } {
    \mathbf{#1}_{#2}
}

\DeclareDocumentCommand \Ts
{  m } {%
T
}

\DeclareDocumentCommand \ones
{ m } {%
	\mathbf{1}_{#1}%
}

\DeclareDocumentCommand \eye
{  O{} } {%
	\mathbf{I}_{#1}%
}
\DeclareDocumentCommand \zeros
{ o o } {
	\mathbf{0}_{		
		\IfNoValueTF{#1}{}{#1}
		\IfNoValueTF{#2}{}{\times#2}		
	}
}
\DeclareDocumentCommand \starprod
{ m m o o}{\IfNoValueTF{#3}{\left(}{#3}{#1} \star {#2}\IfNoValueTF{#4}{\right)}{#4}}
\DeclareDocumentCommand \fu
{ o o }{{\color{red}\textbf{HERE}}}

\DeclareDocumentCommand \fl
{ o o }{
	\mathcal{F}_l%
    \IfNoValueTF{#1}{}{%
    	\IfNoValueTF{#2}{}{%
            \left({#1},{#2}\right)%
        }%
    }%
}

\DeclareDocumentCommand \muDD
{ o }{
	\mu_
	{\IfNoValueTF{#1}{\mf{\Delta}}{#1}}
}
\DeclareDocumentCommand \muD
{ o o }{
	\muDD[#2]	
    \IfNoValueTF{#1}{}{
	\left(#1\right)}
}

\newcommand{\diag}[1]{\text{diag}\left(#1\right)}

\usepackage{tikz}
\usetikzlibrary{shapes.geometric,calc}
\DeclareDocumentCommand \circled
{ m o }{
    \tikz[baseline=(char.base)]{
        \node[shape=circle,fill=white,draw,inner sep=\IfNoValueTF{#2}{2pt}{#2}] (char) {#1};
    }
}

\definecolor{colorA1}{RGB}{10,22,221}
\definecolor{colorA2}{RGB}{153,0,0}
\definecolor{colorL}{RGB}{26,0,128}

\DeclareDocumentCommand \diff
{ o } {
	\,d#1
}

\DeclareDocumentCommand{\crossF}{m o}{\IfValueTF{#2}{#2}{\left}[#1\IfValueTF{#2}{#2}{\right}]^\times}

\title{Optimal Science-time Reorientation Policy for the Comet Interceptor Flyby via Sequential Convex Programming}

\author{Valentin Preda \footnote{Control Systems Engineer, GNC, AOCS \& Pointing Division, ESA/ESTEC.}, Andrew Hyslop \footnote{AOCS Engineer, Vitrociset Belgium S.R.L. - a Leonardo Company - for ESA/ESTEC.} and Samir Bennani \footnote{Senior Advisor, GNC, AOCS \& Pointing Division, ESA/ESTEC.}}
\affil{European Space Agency, ESTEC, 2201 AZ Noordwijk, The Netherlands}

\begin{document}

\maketitle

\begin{abstract}
This paper introduces an algorithm to perform optimal reorientation of a spacecraft during a high speed flyby mission that maximizes the time a certain target is kept within the field of view of scientific instruments. The method directly handles the nonlinear dynamics of the spacecraft, sun exclusion constraint, torque and momentum limits on the reaction wheels as well as potential faults in these actuators. A sequential convex programming approach was used to reformulate non-convex pointing objectives and other constraints in terms of a series of novel convex cardinality minimization problems. These subproblems were then efficiently solved even on limited hardware resources using convex programming solvers implementing second-order conic constraints. The proposed method was applied to a scenario that involved maximizing the science time for the upcoming Comet Interceptor flyby mission developed by the European Space Agency. Extensive simulation results demonstrate the capability of the approach to generate viable trajectories even in the presence of reaction wheel failures or prior dust particle impacts.
\end{abstract}

\section{Introduction}

\subsection{Background and motivation}
Comet Interceptor \cite{Snodgrass2019,Schwamb2020} is an upcoming mission developed on a rapid time scale by the European Space Agency and expected to launch in 2028 towards the Sun-Earth L2 Lagrange point. The mission takes advantage of the extra space available on the launcher of the ARIEL mission and will wait in hibernation until a suitable comet target is found. Ideally, Comet Interceptor will be the first mission to visit a truly pristine comet or an interstellar object visiting the inner Solar System for the first time. The spacecraft is equipped with a multitude of imaging equipment including additional smaller and detachable probes, providing multiple simultaneous sensing of the comet nucleus and its environment. The primary spacecraft will perform a fly-by at \SI{\sim 1000}{km} from the target with a relative velocity between \SIrange[range-phrase={~to~}]{10}{70}{km/s}. %
In terms of attitude guidance and control, this one-shot mission poses a number of challenges, namely the fact that the spacecraft (for some of the study design options) needs to perform an agile and large angle slew to maintain the comet within the field of view during the flyby while also dealing with possible dust particle impacts. The task is further complicated when considering saturations and potential faults in the reaction wheel attitude control system or the sun exclusion constraints needed protect the sensitive imagining equipment. The algorithms also need to be predictive in nature since the optimization trade-offs should be balanced across the whole flyby interval. These requirements motivated the investigation of advanced methods capable of quickly generating viable spacecraft reorientation trajectories subject to a large number of mission constraints. The ability to regenerate these trajectories directly on-board the spacecraft without the need of the ground segment could prove critical in maximizing science time during the very narrow flyby window.

\subsection{Related work}
The constrained attitude guidance problem has received considerable attention in the literature.
Some path planning algorithms work by discretizing the rotational space into a connected graph and then finding a suitable reorientation path by means of a graph search method such as $A^*$ \cite{Kjellberg2013,Tanygin2017} or other random search heuristics \cite{Feron2001,Cui2007}. 
A limitation of these path planning approaches is the spacecraft dynamics is neglected. Therefore, the algorithms do not take advantage of gyroscopic coupling or of the momentum stored in the reaction wheel assembly to perform maneuvers that are more agile.%

Other techniques involve modifying a classical feedback controller with an auxiliary term in order to enforce different types of constraints. For example, the Cassini spacecraft \cite{Singh1997} used a constraint monitor and an avoidance function to ensure sun exclusion constraints on some of its instruments. Potential field methods such as those presented in \cite{Hu2020,Avanzini2009,Radice2007OnConstraints} rely on a repulsive input to steer the closed loop system trajectory away from the constraints. While effective for some simple motion planning problems, is it rather difficult to construct the appropriate control functions that impose more complex constraints or objectives. Furthermore, potential field methods can get trapped in certain local minima and exhibit oscillatory behavior \cite{Koren1991PotentialNavigation}. However, one the most important limitation of these techniques is that they are fundamentally reactive in nature and lack predictive capabilities. This makes them a sub-optimal choice when the objective involves a constrained optimization across a long future horizon, as is the case in the flyby problem outlined in this work.

To fill in this gap, methods relying on real-time convex optimization have been developed for a large number of aerospace applications \cite{Liu2017,Dueri2018,Eren2017} including many that involve non-convex constraints or objectives. These methods leverage the increasing computational capabilities of embedded systems to solve a series of tractable convex optimization problems directly on-board using polynominal time-algorithms. Example applications include pinpoint landers and powered descent vehicles \cite{Pinson2018,Harris2014,Acikmese2013,Malyuta2019}, trajectory planning for space rendezvous and proximity operations \cite{Virgili-Llop2019,Lu2013c} or autonomous spacecraft swarms \cite{Morgan2016}. Heuristic techniques based on the sequential convex programming (SCP) were used in several of these aerospace studies \cite{Mao2016,Bonalli2019,Virgili-Llop2019,Szmuk2018} as a means of handling the non-convex constraints that naturally arise in these applications. These SCP methods were also adapted for the current study, particularly the prior work performed by \cite{Kim2004,UnsikLee2012,Virgili-Llop2019} for the convex parametrization of attitude constrained zones and \cite{Szmuk2018,Malyuta2019} for the discretization techniques. The work of \cite{Kim2004}, was also extended in \cite{Tam2016} to handle logical combinations of multiple keep-in convex pointing constraints by means of mixed integer convex programming (MICP). The MICP method was also used in \cite{Eren2015AGuidance} for constrained attitude guidance. Unfortunately, the solution complexity of the MICP formulation increases exponentially with the number of these binary variables. Furthermore, if the main non-convex optimization heuristic relies on SCP, then this computational cost is compounded by the iterative nature of SCP approach which would require solving a MICP multiple times for each trajectory. To overcome some of these challenges, the authors of \cite{Malyuta2020FastConstraints}, reformulate the mixed-integer problem in terms of state-triggered constraints that can be easily introduced in an SCP loop. 

\subsection{Contributions and paper organization}
The paper focuses on generating spacecraft reorientation trajectories during high speed flyby missions that provide optimal scientific observation time subject to a multitude of constraints and potential faults. To this end, the primary contribution of this study is a novel reformulation of these flyby objectives as a non-convex cardinality minimization problem. In contrast to other works on the topic of constrained attitude guidance, the method relies on an accurate discretization strategy to obtain a trajectory that is dynamically accurate across a large prediction horizon corresponding to the flyby. The method directly takes into account the dynamics of the spacecraft actuated by a reaction wheel assembly as well as the various pointing constraints. Detailed step-by-step instructions are provided to explain how this challenging optimization can be rendered tractable even on limited hardware resources using a sequential convex programming approach. A comprehensive Monte Carlo analysis is used to assess the convergence properties of the algorithm and the quality of the generated trajectories for the Comet Interceptor mission in faulty scenarios and considering prior dust particle impacts.

The rest of the paper is organized in the following way. In \cref{sec:problem_formulation}, the necessary mathematical background is presented and the issue of optimal pointing during flybys is formulated as nonlinear and non-convex optimization problem. The various steps and heuristics used to approach this problem are subsequently presented. \Cref{sec:results} presents and discusses the numerical results obtained during the Monte Carlo campaign. Finally, \cref{sec:conclusions} concludes the paper and outlines possible future extensions of this work.

\section{Problem formulation}
\label{sec:problem_formulation}
\subsection{Overview of guidance objectives}
\begin{figure*}
\centering
\def\svgwidth{0.4\textwidth}
{
\footnotesize
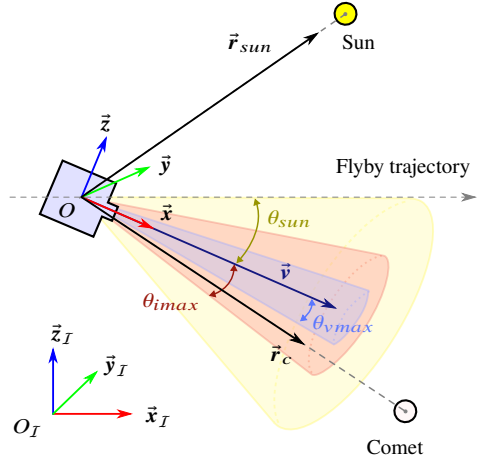
}
\caption{Illustration of the instrument pointing $\sf{\vec{v}}$, comet pointing $\sf{\vec{r}}[c]$ and sun pointing $\sf{\vec{r}}[sun]$ unit vectors as well as the various field of view constraints for the Comet Interceptor mission. Note: In this diagram $\sf{\vec{v}}$ is assumed to be aligned with the $\sf{\vec{x}}$ body axis.}
\label{fig:frames}
\end{figure*}

\Cref{fig:frames} illustrates the flyby scenario of the Comet Interceptor mission together with different unit vectors used to describe the spacecraft state. The vector $\sf{\vec{v}}$
is aligned along the sensor's boresight direction and has the known static coordinates $\sf{v} \in \mathbb{R}^3$ in the spacecraft body frame $\mathcal{F}_\mathcal{B} := \coordframe{O}{\sf{\vec{x}}}{\sf{\vec{y}}}{\sf{\vec{z}}}$. The vector $\sf{\vec{r}}[c]$ points towards the target comet and $\sf{\vec{r}}[sun]$ towards the sun. It is assumed that an on-board navigation subsystem provides the coordinates $\sf{r}[c](t), ~\sf{r}[sun](t) \in \mathbb{R}^3$ of these vectors in an arbitrary inertial coordinate frame $\mathcal{F}_\mathcal{I} := \coordframe{O_\mathcal{I}}{\sf{\vec{x}}[\mathcal{I}]}{\sf{\vec{y}}[\mathcal{I}]}{\sf{\vec{z}}[\mathcal{I}]}$. To conduct scientific observations, the angle between the $\sf{\vec{v}}$ and $\sf{\vec{r}}[c]$ vectors must be kept below the maximum half angle of field of view angles $\theta_{vmax}$ for the visual camera and $\theta_{imax}$ for the infrared camera. The angle between $\sf{\vec{v}}$ and the sun pointing vector $\sf{\vec{r}}[sun]$ needs to be maintained above the $\theta_{sun}$ limit to protect the sensitive instruments.
The goal of the guidance optimization problem is to compute an optimal reaction wheel control signal that maximizes the time the target is kept within the field of view of the scientific instruments during a flyby interval $t \in [0, t_f]$. Using auxiliary reaction control thrusters during the fly-by is not allowed as they may contaminate some of the science instruments.
\begin{remark}
An alternative design of the Comet Interceptor spacecraft incorporates a one degree of freedom scanning mechanism that can help track the comet during the flyby. A separate tracking controller for this mechanism would then center the comet on the imaging sensor once the target is within a certain field of view cone. For brevity, this alternative design is not discussed in this paper but the proposed guidance method can be directly applied to cover it by simply expanding the field of view of limits $\theta_{vmax}$ and $\theta_{imax}$ in one axis to account for the increased authority provided by the new scan mechanism. However, it is likely that the alternate design would also be associated with a fixed guidance attitude due to the presence of a dust shield on just one spacecraft face. In this scenario the convex optimizer may only prove useful in a contingency event where a large dust particle impact causes significant off-pointing and wheel saturation upon recovery. In such a case a real-time attitude guidance optimization could be performed to minimize science outage given the on-board conditions.
\end{remark}
\subsection{Equations of motion and system dynamics}
\label{sec:eq_motion}
Let the vector $\sf{\omega}(t) \in \mathbb{R}^3$ denote the spacecraft angular velocity in the body frame $\mathcal{F}_\mathcal{B}$ at time $t$ while the unit quaternion $\sf{q}(t) \in \mathbb{R}^4$ with $\sf{q}^\ts\sf{q} = 1$ is used to rotate from the inertial frame $\mathcal{F}_\mathcal{I}$ to $\mathcal{F}_\mathcal{B}$. The matrix representation of the quaternion is defined as: $\sf{q}(t) = \iv{\sf{q}[v]^\ts & q_s}{}^\ts$ where $q_s \in \mathbb{R}$ denotes the scalar part and $\sf{q}[v] \in \mathbb{R}^3$ denotes the vector part. Considering that the spacecraft is actuated by a set of $n_w$ reaction wheels, its attitude kinematics and dynamics are described by:
\begin{equation}
\begin{aligned}
    \sf{\dot{q}}(t) &= \frac{1}{2}\sf{q}(t) \otimes \iv{\sf{\omega(t)} \\ 0}\\
    \sf{J}\sf{\dot{\omega}}(t) &= \big[\sf{J} \sf{\omega}(t) + \sf{L}\sf{h}(t)\big] \times \sf{\omega}(t) -\sf{L}\sf{\tau}(t) + \sf{d}(t)\\
    \sf{\dot{h}}(t) &= \sf{\tau}(t)
\end{aligned}
\label{eq:dynamics}
\end{equation}
where the operator $\otimes$ denotes the quaternion product, $\sf{J} \in \mathbb{R}^{3 \times 3}$ is moment of inertia in the body frame, $\sf{h}(t) \in \mathbb{R}^{n_w}$ is vector containing the angular momentum stored in each of $n_w$ wheels, $\sf{\tau}(t) \in \mathbb{R}^{n_w}$ is the motor torque applied to the wheels, $\sf{L} \in \mathbb{R}^{3 \times n_w}$ is reaction wheel torque distribution matrix and $\sf{d}(t) \in \mathbb{R}^3$ is a disturbance torque. 
\begin{remark}
For the trajectory generation algorithm proposed in this paper, the spacecraft was assumed to be rigid with perfectly known states and the perturbations $\sf{d}$ were assumed to be null during the flyby. In reality, the spacecraft structure can include flexible elements, its states are estimated with residual noise and disturbances are introduced from both external sources and various on-board equipment. An additional control system would need to be developed in order to track the reference trajectory provided by the guidance algorithm and reject additional disturbances. Such a control system design was beyond the scope of this paper.
\end{remark}
To avoid numerical issues the different states and inputs of the nonlinear dynamical system \cref{eq:dynamics} have been scaled with respect to their maximum expected magnitudes. In a compact notation, this newly scaled dynamical system is written as:
\begin{equation}
    \dot{\sf{x}}(t) = \sf{f}\left(\sf{x}(t), \sf{u}(t)\right)\quad\text{where}\quad
    \begin{aligned}
    \vf{x}(t) &= \iv{\sf{q}(t) \\ \sf{W}[\omega]\sf{\omega}(t) \\ \sf{W}[h]\sf{h}(t)} \in \mathbb{R}^{7 + n_w}
    \\
    \sf{u}(t) &= \sf{W}[\tau] \sf{\tau}(t) \in \mathbb{R}^{n_w}
    \end{aligned}
    \quad\text{and}\quad
    {\footnotesize
    \begin{aligned}
        \sf{W}[\omega] &= \diag{\sf{\omega}[max]}^{-1} \\
        \sf{W}[h] &= \diag{\sf{h}[max]}^{-1} \\
        \sf{W}[\tau] &= \diag{\sf{\tau}[max]}^{-1}
    \end{aligned}
    }
    \label{eq:dynamics_compact}
\end{equation}
where $\sf{x}$ and $\sf{u}$ represent the newly scaled state vector and control inputs, $\sf{\tau}[max],~\sf{h}[max] \in \mathbb{R}^{n_w}$  represent the maximum motor torques and angular momenta for each of wheels and $\sf{\omega}[max] \in \mathbb{R}^3$ the maximum angular rates of the spacecraft. 

\subsection{Convex pointing constraints}
To express the guidance problem in a more rigorous manner, it is useful to first the consider line of sight constraints.
Let $\sf{r} \in \mathbb{R}^3$ be coordinates in the inertial frame of a target pointing unit vector $\vec{\sf{r}}$ and $\sf{\nu} \in \mathbb{R}^3$ the body frame coordinates of an instrument pointing vector $\sf{\vec{\nu}}$. The cosine of the angle between these two vectors is given by the dot product relationship \cite{Kim2004,Kim2010,UnsikLee2012,Virgili-Llop2019a}:
\begin{equation}
    \cos{\theta} = \vec{\sf{r}} \cdot \vec{\sf{\nu}} 
    =-\sf{q}^\ts\sf{P}(\sf{r}, \sf{\nu})\sf{q}
    \quad
    \text{with}
    \quad
    \sf{P}(\sf{r}, \sf{\nu}) :=
    \iv{
    \crossF{\sf{r}} & \sf{r} \\
    -\sf{r}^\ts & 0
    }
    \iv{
    -\crossF{\sf{\nu}} & \sf{\nu} \\
    -\sf{\nu}^\ts & 0
    }
    \label{eq:cos_1st_form}
\end{equation}
where the operator $\crossF{\cdot}$ is equal to the left-hand side equivalent of the vector cross product in matrix form, i.e.
$\crossF{\sf{a}} =
    \left[
    \begin{smallmatrix}
    0 & -a_3 & a_2 \\ a_3 & 0 & -a_1 \\ -a_2 & a_1 & 0
    \end{smallmatrix}
    \right]$ for $\sf{a} = \left[\begin{smallmatrix}a_1 & a_2 & a_3\end{smallmatrix}\right]^\ts \in \mathbb{R}^3$. Since both $\sf{r}$ and $\sf{\nu}$ are unit vectors and $\vf{P}$ is a product of two skew-symmetric matrices that commute under multiplication, it follows that $\sf{P}$ is an indefinite matrix with eigenvalues $\pm 1$. Considering that $\sf{q}^\ts\sf{q} = 1$, eq. \cref{eq:cos_1st_form} can be rewritten in terms of semidefinite quadratic forms in two equivalent ways:
\begin{equation}
    \cos{\theta} = 1 - \sf{q}^\ts\left(\eye + \sf{P}\right)\sf{q} = \sf{q}^\ts\left(\eye - \sf{P}\right)\sf{q} - 1
\label{eq:cos_2nd_form}
\end{equation}
    Depending on the type of target, the angle $\theta$ must be kept either below a certain maximum value $\theta_{max}$ or a above a minimum value $\theta_{min}$. In terms of cosines and using \cref{eq:cos_2nd_form}, these constraints can be expressed as:
    \begin{equation}
\begin{alignedat}{3}
    \cos{\theta} &\ge \cos{\theta_{max}} &&\Longleftrightarrow \sf{q}^\ts\left(\eye + \sf{P}\right)\sf{q} &&\le 1-\cos{\theta_{max}}
    \\
    \cos{\theta} &\le \cos{\theta_{min}} &&\Longleftrightarrow \sf{q}^\ts\left(\eye - \sf{P}\right)\sf{q} &&\le 1+\cos{\theta_{min}}
    \label{eq:constraint_cos}
\end{alignedat}
\end{equation}
In this case, the matrices $\eye \pm \sf{P}$ are positive semidefinite with eigenvalues $0$ and $2$. Therefore, both constraints in \cref{eq:constraint_cos} are convex quadratic and can be equivalently described as the following second-order conic constraints: 
\begin{equation}
\begin{alignedat}{3}
    \cos{\theta} &\ge \cos{\theta_{max}} &&\Longleftrightarrow     \|\sf{N}(\sf{r}, \sf{\nu})\sf{q}\|_2 &&\le \sqrt{1 - \cos{\theta_{max}}}
    \\
    \cos{\theta} &\le \cos{\theta_{min}} &&\Longleftrightarrow     \|\sf{M}(\sf{r}, \sf{\nu})\sf{q}\|_2 &&\le \sqrt{1 + \cos{\theta_{min}}}
\end{alignedat}
\quad
\text{with}
\quad
\begin{aligned}
    \eye + \sf{P} = \sf{N}^\ts\sf{N} \\
    \eye - \sf{P} = \sf{M}^\ts\sf{M}
\end{aligned}
\label{eq:constraint_cos_socp}
\end{equation}
where $\sf{N}$ and $\sf{M}$ can be obtained using an eigendecomposition of the matrices $\eye \pm \sf{P}$. Whenever $\|\sf{N}(\sf{r}, \sf{\nu})\sf{q}\|_2 \approx 0$ it follows that $\cos{\theta} \approx 1$ and therefore minimizing this norm drives $\sf{\vec{r}}$ and $\sf{\vec{\nu}}$ towards collinearity.

\subsection{Science objectives as a nonlinear cardinality optimization problem}
The general convex pointing constraints from \cref{eq:constraint_cos_socp} can now be applied to the specific case of guidance problem introduced in \cref{sec:eq_motion} where the end goal is to maximize the total science time. In this case, the constraint related to the solar exclusion angle $\theta_{sun}$ can be readily described as:
\begin{equation}
    \big\|\sf{M}(\sf{r}[sun](t), \sf{v})\sf{q}(t)\big\|_2 \le \sqrt{1 + \cos{\theta_{sun}}}
\end{equation}
Similarly, the pointing constraints related to the imagining cameras are expressed as:
\begin{equation}
    \begin{aligned}
    \big\|\sf{N}(\sf{r}[c](t), \sf{v})\sf{q}(t)\big\|_2 &\le \sqrt{1 - \cos{\theta_{vmax}}}
    + \sf{\gamma}(t); \quad \sf{\gamma}(t) \ge 0\\
    \big\|\sf{N}(\sf{r}[c](t), \sf{v})\sf{q}(t)\big\|_2 &\le \sqrt{1 - \cos{\theta_{imax}}} + \sf{\zeta}(t); \quad \sf{\zeta}(t) \ge 0\\
    \big\|\sf{N}(\sf{r}[c](t), \sf{v})\sf{q}(t)\big\|_2 &\le \sf{\eta}(t)
    \end{aligned}
    \label{eq:fov_constraint_1}
\end{equation}
where the slack variables $\sf{\gamma}(t), \sf{\zeta}(t) \in \mathbb{R}_+$ quantify the magnitude of constraint violation at time $t$ and make the pointing constraints soft. Whenever $\sf{\gamma}$ or $\sf{\zeta}$ are zero, the comet target is within the field of view of the visual or infrared camera. Maximizing visual or infrared science time is therefore equivalent to minimizing the number of nonzero values in either $\sf{\gamma}(t)$ or $\sf{\zeta}(t)$. The overall pointing error between the comet pointing vector $\sf{\vec{r}}[c]$ and instrument vector $\sf{\vec{v}}$ during the flyby time $t \in [0, t_f]$ is a secondary performance objective optimized through a minimization of the values in $\sf{\eta}(t) \in \mathbb{R}_+$. Taking these facts into account, the constrained pointing optimization can be formulated as follows:
\begin{equation}
    \begin{gathered}
    \minimize_{\sf{x},\sf{u},\sf{\gamma},\sf{\xi},\sf{\zeta},\sf{\rho}}\quad
    J_{nl} := 
    \int_{0}^{t_{f}}
    \sf{\beta}\big[\card\big(\sf{\gamma}(t)\big) \enspace \card\big(\sf{\zeta}(t)\big) \enspace \sf{\eta}(t) \enspace \sf{\rho}(t) \Big]^\ts
    \diff[t] \quad \text{ subject to } \\
    \begin{alignedat}{2}
    \sf{\dot{x}}(t) = \sf{f}\left(\sf{x}(t), \sf{u}(t)\right)& &&\quad\text{nonlinear dynamics} \\
    \sf{x}(0) = \sf{x}[init]& &&\quad\text{ initial conditions} \\
     \big\|\sf{M}(\sf{r}[sun](t), \sf{v})\sf{q}(t)\big\|_2 \le \sqrt{1 + \cos{\theta_{sun}}}& &&\quad\text{solar exclusion}\\
    \begin{alignedat}{2}
    \big\|\sf{N}(\sf{r}[c](t), \sf{v})\sf{q}(t)\big\|_2 &\le \sqrt{1 - \cos{\theta_{vmax}}} + \sf{\gamma}(t); \quad &&\sf{\gamma}(t) \ge 0 \\
    \big\|\sf{N}(\sf{r}[c](t), \sf{v})\sf{q}(t)\big\|_2 &\le \sqrt{1 - \cos{\theta_{imax}}} + \sf{\zeta}(t); \quad &&\sf{\zeta}(t) \ge 0
     \end{alignedat}& &&
     \quad
     \begin{aligned}
     &\text{visual field of view} \\
     &\text{infrared field of view}
     \end{aligned}
     \\
     \big\|\sf{N}(\sf{r}[c](t), \sf{v})\sf{q}(t)\big\|_2 \le \sf{\eta}(t)& &&\quad\text{line of sight error}\\
      \|\sf{u}(t)\|_2 \le \sf{\rho}(t)& &&\quad\text{ control energy} \\
    |\sf{\tau}(t)| \preccurlyeq \sf{\tau}[max];
    \quad|\sf{h}(t)| \preccurlyeq \sf{h}[max];
    \quad|\sf{\omega}(t)| \preccurlyeq \sf{\omega}[max] & &&\quad\text{max torque, momentum and angular rates}
    \end{alignedat}
    \end{gathered}
    \label{eq:nonlinear_optimization}
\end{equation}
where the cardinality function $\card(\sf{a})$ denotes number of nonzero elements in a vector $\sf{a}$, the operator $\preccurlyeq$ denotes elementwise inequality, $\sf{x}[init]$ represents the initial state. The vector $\sf{\beta} = \iv{\sf{\beta}[0] & \dots & \sf{\beta}[3]}{}^\ts \in \mathbb{R}^4$ contains the weights that determine the relative importance of each of the terms in the cost function and therefore the trade-off between the different and potentially conflicting performance objectives. The values $\sf{\beta}[0]$ and $\sf{\beta}[1] $ scale the cost of violating the field of view constraint of the visual and infrared camera respectively. $\sf{\beta}[2]$ scales the cost of any nonzero pointing error. $\sf{\beta}[3]$ places a small cost on the energy of the control signal to reduce unnecessary control action. This weight could also be increased if power budget is a serious concern. 

\subsection{Sequential convex programming}
The optimization problem given in \cref{eq:nonlinear_optimization} is challenging to solve directly since it involves the nonlinear dynamics constraint as well as the cardinally function in the objective. To tackle these issues, an SCP technique similar to the ones presented in \cite{Bonalli2019,Mao2016,Szmuk2018,Virgili-Llop2019a} will be employed in this study. The core idea of these methods is to replace each of nonconvex terms in the original nonlinear optimization problem by a convex approximation that is obtained following a linearization around a previous solution. The resulting linear problem is then discretized and efficiently solved using convex optimization tools. The solution is then used as a new linearization trajectory for a subsequent iteration of the algorithm and these steps are repeated until convergence. Details about all these steps in the algorithm are presented and discussed in the following sections.
\subsubsection{Linearization}
To eliminate the challenging nonlinear dynamics constraint \cref{eq:dynamics} from the optimization problem \cref{eq:nonlinear_optimization}, the following first-order Taylor expansion around a previous trajectory $\sf{\overline{x}}(t), \sf{\overline{u}}(t)$ is performed:
\begin{equation}
\begin{gathered}
    \dot{\sf{x}}(t) = \sf{f}\left(\sf{x}(t), \sf{u}(t)\right) \approx \sf{A}(t)\sf{x}(t) + \sf{B}\sf{u}(t) + \sf{s}(t) \quad\text{where}\\
    \sf{A}(t) := \nabla_{\sf{x}}\sf{f}(\sf{\overline{x}}(t))
    =
    \iv{
    \dfrac{1}{2}
    \left[
    \begin{array}{cc}
    -\crossF{\sf{\overline{\omega}}(t)}[\big] & \sf{\overline{\omega}}(t) \\ -\sf{\overline{\omega}}(t)^\ts & 0
    \end{array}
    \right]
    &
    \dfrac{1}{2}
    \left[
    \begin{array}{c}
    \sf{\overline{q}}[s](t)\eye[3] + \crossF{\sf{\overline{q}}[v](t)}\\
    -\sf{\overline{q}}[v](t)^\ts
    \end{array}
    \right]\sf{W}[\omega]^{-1}
    & \zeros[4][n_w] \\
    \zeros[3][4] & \sf{J}^{-1}\left(\crossF{\sf{J}\sf{\overline{\omega}}(t)}[\big]- \crossF{\sf{\overline{\omega}}(t)}[\big]\sf{J} + \crossF{\sf{L}\sf{\overline{h}}(t)}[\big]\right)~ & -\sf{W}[\omega]\sf{J}^{-1}\crossF{\sf{\overline{\omega}}(t)}[\big]\sf{L}\sf{W}[h]^{-1} \\
    \zeros[n_w][4] & \zeros[n_w][3] & \zeros[n_w]
    }
    , \\
    \sf{B} := \nabla_{\sf{u}}\sf{f} = \iv{
    \zeros[4][n_w] \\
    -\sf{W}[\omega]\sf{J}^{-1}\sf{L}\sf{W}[u]^{-1} \\
    \sf{W}[h]\sf{W}[u]^{-1}
    },
    \quad
    \sf{s}(t) = -\sf{A}(t)\sf{\overline{x}}(t) - \sf{B}\sf{\overline{u}}(t),
    \quad
    \vf{\overline{x}}(t) = \iv{\sf{\overline{q}}(t) \\ \sf{W}[\omega]\sf{\overline{\omega}}(t) \\ \sf{W}[h]\sf{\overline{h}}(t)}
\end{gathered}
\label{eq:linearized}
\end{equation}
The notation $\nabla_{\sf{x}}\sf{f}$ indicates the Jacobian of a function $\sf{f}$ with respect to a vector $\sf{x}$. The new 
linear time-varying constraint only provides a good approximation within a narrow corridor around the previous trajectory. Therefore, new trust region constraints will be introduced to the original optimization problem to keep the solution in a region where this approximation is accurate enough. A similar linearization was used to eliminate the non-convex cardinality terms from the optimization objective by means of an weighted iterated $l_1$ heuristic \cite{Boyd2007}, i.e.
\begin{equation}
\card\big(\sf{\gamma}(t)\big) \approx \frac{\sf{\gamma}(t)}{\varepsilon + \sf{\overline{\gamma}}(t)}
\label{eq:card_approx}
\end{equation}
where $\sf{\overline{\gamma}}(t) \in \mathbb{R}_+$ denotes the solution from the previous iteration and $\varepsilon = 10^{-3}$ is a small term introduced to avoid division by zero whenever $\sf{\overline{\gamma}}(t) = 0$. The linearization of $\card\big(\sf{\zeta}(t)\big)$ is done analogously.
\subsubsection{Discretization}
The newly linearized continuous-time optimization problem can now be rendered tractable by numerical tools by means of discretization. Following a similar notation and approach as in \cite{Malyuta2019,Szmuk2018}, the time interval $t \in [0, t_f]$ is subdivided into a grid of $N$ sampling times $t_k$ with $k \in \{0, \dots, N-1\}$ such that $t_0 =0$ and $t_{N-1}=t_f$. The control signal within each interval is assumed to be linearly interpolated between the discrete values $\sf{u}[k]$ and $\sf{u}[k+1]$, i.e.
\begin{equation}
\begin{gathered}
    \sf{u}(t) = \lambda_k^-(t)\sf{u}[k] + \lambda_k^+(t)\sf{u}[k+1],\quad\forall t\in(t_k, t_{k+1}]
\end{gathered}
\label{eq:foh}
\end{equation}
where  $\lambda_k^-(t) = (t_{k+1} - t)/(t_{k+1} - t_k)$ and $\lambda_k^+(t) = 1 - \lambda_k^-(t)$. For each interval, the state trajectory $\sf{x}(t)$ satisfying the linearized dynamics \cref{eq:linearized} is
\begin{equation}
    \sf{x}(t) = \sf{\Phi}(t, t_k)\sf{x}(t_k) + 
    \int_{t_k}^t\sf{\Phi}(t, \xi)\sf{B}\sf{u}(\xi)d\xi + \int_{t_k}^t\sf{\Phi}(t, \xi)\sf{s}(\xi) \diff[\xi]
    \quad
    \begin{aligned}
    \forall t \in (t_k, t_{k+1}]
    \end{aligned}
    \label{eq:state_propagation}
\end{equation}
where $\sf{\Phi}(\xi, t_k)$ is the following state transition matrix from time $t_k$ to time $\xi$:
\begin{equation}
    \sf{\Phi}(\xi, t_k) = \eye + \int_{t_k}^{\xi}\sf{A}(t)\sf{\Phi}(t, t_k) \diff[\tau]
    \label{eq:phi}
\end{equation}
Using \cref{eq:foh} and the fact that $\sf{\Phi}(t_{k+1}, \xi) = \sf{\Phi}(t_{k+1},t_k)\sf{\Phi}(\xi, t_k)^{-1}$, the discrete LTV dynamics for each subinterval can be obtained based on a previous discrete linearization trajectory $\sf{\overline{x}}[k], \sf{\overline{u}}[k]$ with $k \in \{0, \dots, N-1\}$ as:
\begin{gather}
\begin{gathered}
    \sf{x}[k+1] = \sf{A}[k]\sf{x}[k] + \sf{B}[k]^-\sf{u}[k] + \sf{B}[k]^+\sf{u}[k + 1] + \sf{s}[k] \quad\text{with}\\
    \sf{A}[k] = \sf{\Phi}(t_{k+1}, t_k) = \eye + \int_{t_k}^{t_{k+1}}\sf{A}(\xi)\sf{\Phi}(\xi, t_k) \diff[\xi]
    ,\\
    \sf{B}[k]^- = \sf{A}[k]\int_{t_k}^{t_{k+1}}\sf{\Phi}(\xi, t_{k})^{-1}\lambda_k^{-}(\xi)\sf{B} \diff[\xi],\\
    \sf{B}[k]^+ = \sf{A}[k]\int_{t_k}^{t_{k+1}}\sf{\Phi}(\xi, t_{k})^{-1}\lambda_k^{+}(\xi)\sf{B} \diff[\xi], \\
    \sf{s}[k] = \sf{A}[k]\int_{t_k}^{t_{k+1}}\sf{\Phi}(\xi, t_{k})^{-1}\underbrace{\left(-\sf{A}(\xi)\sf{\overline{x}}(\xi) - \sf{B}\sf{\overline{u}}(\xi)\right)}_{\sf{s}(\xi)} \diff[\xi]  \quad\text{where}
    \label{eq:discretization}
\end{gathered}
\\  \sf{\overline{x}}(\xi) =  \sf{\overline{x}}[k] + \int_{t_k}^{\xi} \sf{f}\big(\sf{\overline{x}}(t), \sf{\overline{u}}(t)\big)\diff[t]\quad\text{and}\quad
    \sf{\overline{u}}(\xi) = \lambda_k^-(\xi)\sf{\overline{u}}[k] + \lambda_k^+(\xi)\sf{\overline{u}}[k+1]\quad
    \forall \xi \in \left(t_k, t_k+1\right]
    \label{eq:discretization_state_continuous}
\end{gather}
\begin{remark}
The precise values of $\sf{A}[k]$, $\sf{B}[k]^-$, $\sf{B}[k]^+$ and $\sf{s}[k]$ can be calculated to arbitrary precision and in parallel for each subinterval. A more precise integration leads to better discretization accuracy and potential improvements in the convergence rate of the overall algorithm at the cost of more computational effort for each discretization step \cite{Malyuta2019}. Within each subinterval, the integrals are computed simultaneously and the most expensive operation is the inversion of the state transition matrix $\sf{\Phi}(\xi, t_{k})^{-1}$ performed using LU decomposition.
\end{remark}
\subsubsection{Convex form of the optimization problem}
Following the previous guidelines, the original nonlinear optimization problem from \cref{eq:nonlinear_optimization} was approximated by the following convex form:
\begin{equation}
    \begin{gathered}
    \minimize_{\substack{\sf{x}[k],\sf{u}[k],\sf{\gamma}[k],\sf{\zeta}[k],\sf{\eta}[k],\sf{\rho}[k],\sf{\delta}[x,k],\sf{\delta}[u,k]\\k=\{0, \dots, N-1\}}}
    \quad
    J :=
    \sum_{k=0}^{N-1}
    \sf{\beta}\left[\frac{\sf{\gamma}[k]}{\varepsilon + \sf{\overline{\gamma}}[k]} \quad
    \frac{\sf{\zeta}[k]}{\varepsilon + \sf{\overline{\zeta}}[k]} \quad \sf{\eta}[k] \quad \sf{\rho}[k]\quad \sf{\delta}[x,k]\quad
    \sf{\delta}[u,k]\right]^\ts
    \quad\text{ subject to } \\
    \begin{alignedat}{2}
    \sf{x}[k+1] = \sf{A}[k]\sf{x}[k] + \sf{B}[k]^-\sf{u}[k] + \sf{B}[k]^+\sf{u}[k + 1] + \sf{s}[k]& &&\quad
    \text{dynamics linearized around }\sf{\overline{x}}, \sf{\overline{u}}\text{ for } k \in \{0,\, ...\, , N - 2\}
    \\
    \sf{x}[0] = \sf{x}[init]& &&\quad\text{initial conditions} \\
    \text{ for } k \in \{0,\, ...\, , N - 1\}: \\
     \big\|\sf{M}(\sf{r}[sun](t_k), \sf{v})\sf{q}[k]\big\|_2 \le \sqrt{1 + \cos{\theta_{sun}}}& &&\quad\text{solar exclusion}\\
    \begin{alignedat}{2}
    \| \sf{N}(\sf{r}[c](t_k), \sf{\nu})\sf{q}[k]\|_2 &\le \sqrt{1 - \cos{\theta_{vmax}}} + \sf{\gamma}[k]; \quad &&\sf{\gamma}[k] \ge 0 \\
     \| \sf{N}(\sf{r}[c](t_k), \sf{\nu})\sf{q}[k]\|_2 &\le \sqrt{1 - \cos{\theta_{imax}}} + \sf{\zeta}[k]; \quad &&\sf{\zeta}[k] \ge 0
     \end{alignedat}& &&
     \quad
     \begin{aligned}
     &\text{visual field of view} \\
     &\text{infrared field of view}
     \end{aligned}
    \\
    \| \sf{N}(\sf{r}[c](t_k), \sf{\nu})\sf{q}[k]\|_2 \le \sf{\eta}[k]& &&\quad\text{line of sight error}\\
    \|\sf{u}[k]\|_2 \le \sf{\rho}[k]& &&\quad\text{ control energy} \\
    |\sf{\tau}[k]| \preccurlyeq \sf{\tau}[max];
    \quad|\sf{h}[k]| \preccurlyeq \sf{h}[max];
    \quad|\sf{\omega}[k]| \preccurlyeq \sf{\omega}[max] & &&\quad\text{max torque, momentum and angular rates} \\
    \|\sf{\overline{x}}[k]-\sf{x}[k]\|_2 \le \sf{\delta}[x,k];\quad  \|\sf{\overline{u}}[k]-\sf{u}[k]\|_2 \le \sf{\delta}[u,k]& &&\quad\text{state and torque deviations}\\
    \sf{\delta}[x,k] \le \delta_{xmax};\quad \sf{\delta}[u,k] \le \delta_{umax}& &&\quad\text{max trust region sizes}
    \end{alignedat}
    \end{gathered}
    \label{eq:convex_optimization}
\end{equation}
The new problem contains extra trust region constraints on the control input and state trajectories in addition to the linearized discrete-time versions of the constraints and objectives from \cref{eq:nonlinear_optimization}.  The extra terms $\sf{\delta}[x]$ and $\sf{\delta}[u]$ in the objective introduce penalties on deviations from the previous state and control input trajectories and helps steer the iterative algorithm towards convergence. The weight of the penalty in the cost function is controlled by the new scalars $\sf{\beta}[4], \sf{\beta}[5]$ in the scaling vector $\sf{\beta} = \iv{\sf{\beta}[0] & \dots & \sf{\beta}[5]}{}^\ts \in \mathbb{R}^6$. The maximum deviations are also limited to $\delta_{xmax}$ for the state trajectories and $\delta_{umax}$ for the control inputs. 
\begin{remark}
In this specific form, the convex problem \cref{eq:convex_optimization} can ensure constraint satisfaction only at the $N$ sampling times $t_k$. If the temporal grid is too sparse, the constraints related to the pointing objectives, maximum wheel momentum or spacecraft angular rates could be violated during the intersample period. More sophisticated methods such as the one presented in \cite{Dueri2017} could be employed to guarantee that the conditions hold even in the intersample time. For this paper however, the motor torque within each subinterval is linearly interpolated between the endpoints. As such, the maximum intersample violations of all these constraints stay bounded and decrease in magnitude with every increase in $N$. To guarantee proper intersample behavior, the simpler method adopted in this study was to slightly tighten the constraints applied at the sampling times. More precisely, the values of $\sf{h}[max], \sf{\omega}[max], \theta_{vmax}$ and $\theta_{imax}$ were all decreased by 3\% prior to their usage in \cref{eq:convex_optimization}.
\end{remark}

\subsubsection{Trust region update policy}
If the solution is allowed to deviate too far from the previous trajectory, the linear approximation \cref{eq:linearized} might be inaccurate and possibly lead to a solution that is not dynamically feasible. Propagating this unfeasible solution into subsequent iterations as the new linearization trajectory can lead to convergence issues. On the other hand, if the trust regions are heavily constrained, progress is slow and the end solution might be very far from optimal since exploration is discouraged. A heuristic method that achieves a good balance involves modifying the limits $\delta_{xmax}$, $\delta_{umax}$ at each iteration based on metric evaluating the quality of the new solution \cite{Conn2000,Mao2016}. For this application, the chosen metric is:
\begin{equation}
    \epsilon_x = \sum_{k=0}^{N-1}\left\|\sf{x}[k]^\star - \sf{x}[k]\right\|_2
    \label{eq:epsilon}
\end{equation}
where $\sf{x}[k]$ is the state at time $t_k \in [0, t_f]$ obtained using the convex optimization \cref{eq:convex_optimization} and $\sf{x}[k]^\star$ is the true state at time $t_k$. This true state trajectory is found by integrating the nonlinear dynamics equation \cref{eq:dynamics} using the control solutions $\sf{u}[k]$ resulting from  \cref{eq:convex_optimization}. Whenever $\epsilon_x$ exceeds a certain threshold $\epsilon_{max}$, the solution is discarded as unfeasible, the maximum trust region sizes $\delta_{xmax}$, $\delta_{umax}$ are reduced by multiplying them with a factor $\kappa^{-} \in (0, 1]$ and convex optimization is re-executed. Otherwise, if $\epsilon_x < \epsilon_{max}$, the solution is accepted and the trust region sizes are expanded by multiplying them with $\kappa^{+} \ge 1$.
\begin{remark}
The update policy adopted in this study is one among many other possibilities. This law together with the parameter values $\kappa^{-}$, $\kappa^{+}$ and $\epsilon_x$ have a large impact on the convergence performance. A detailed comparatives analysis of the different update policies may be necessary, however such an analysis is beyond the scope this paper.
\end{remark}

\subsubsection{Transcription to a canonical convex optimization form}
The optimization problem \cref{eq:convex_optimization} can be efficiently tackled by a variety of convex solvers implementing second order conic constraints. These multiple solvers accept the problem description in different canonical forms. Automatic transcription tools such as \cite{Agrawal2018} enable users to describe and solve convex optimization problem in a higher level of abstraction without the need of manually converting to the canonical form of the underlying solver. While extremely useful for prototyping ideas, these tools are still inferior to the manual assembly of the canonical form when the optimization problem depends on a large number of parameters and real-time performance is required. In the case of this paper, the open source solver ECOS \cite{domahidi2013ecos} was adopted and \cref{eq:convex_optimization} was assembled in the following standard form:
\begin{equation}
\begin{gathered}
    \minimize_{\sf{x}[s]}
    \quad
    \sf{c}[s]^\ts \sf{x}[s]
    \quad\text{ subject to } \\
    \sf{A}[s]\sf{x}[s] = \sf{b}[s]\\
    \sf{h}[s] - \sf{G}[s]\sf{x}[s] \in \mathbb{R}_+^{m_{l}} \times \mathbb{Q}^{n_1} \times \dots \times \mathbb{Q}^{n_{m}}
    \label{eq:ecos_form}
\end{gathered}
\quad\text{where}\quad
\begin{gathered}
\sf{x}[s] = \iv{\sf{x}[s,0]^\ts & \dots & \sf{x}[s,N-1]^\ts}{}^\ts \\
\sf{x}[s,k] = \iv{\sf{x}[k]^\ts & \sf{u}[k]^\ts & \sf{\gamma}[k] & \sf{\zeta}[k] & \sf{\eta}[k] & \sf{\rho}[k] & \sf{\delta}[x,k] & \sf{\delta}[u,k]}^\ts \\
\sf{c}[s] = \iv{\sf{c}[s, 0]^\ts & \dots & \sf{c}[s, N-1]^\ts}^\ts \\
\sf{c}[s,k] = \iv{\zeros[1][7 + 2 n_w] & \dfrac{\sf{\beta}[0]}{\varepsilon + \sf{\overline{\gamma}}[k]} & \dfrac{\sf{\beta}[1]}{\varepsilon + \sf{\overline{\zeta}}[k]} & \sf{\beta}[2] & \dots & \sf{\beta}[5]}^\ts\\
k = \{0,\dots,N-1\}
\end{gathered}
\end{equation}
and $\mathbb{Q}^n$ denotes the $n$-dimensional second order cone, i.e. $\mathbb{Q}^n := \left\{\left.\left[\SmallMatrix{t \\ \sf{b}}\right] \in \mathbb{R}^n \;\right\rvert\;\sf{b} \in \mathbb{R}^{n-1},  \|\sf{b}\|_2 \le t \right\}$. Let $\sf{\mathcal{Z}}[\sf{a}]$ denote a selection matrix that extracts a particular signal $\sf{a}$ from the vector of variables $\sf{x}[s]$ (for example $\sf{u}[0] = \sf{\mathcal{Z}}[\sf{u}[0]]\sf{x}[s]$). In this case, the equality constraints from \cref{eq:convex_optimization} can be expressed in the canonical form as:
\begin{equation}
    \underbrace{
    \iv{
    \sf{\mathcal{Z}}[\sf{x}[0]] \\
    \sf{\mathcal{Z}}[\sf{x}[1]] - \sf{A}[0]\sf{\mathcal{Z}}[\sf{x}[0]] - \sf{B}[0]^-\sf{\mathcal{Z}}[\sf{u}[0]]- \sf{B}[0]^+\sf{\mathcal{Z}}[\sf{u}[1]]\\
    \vdots \\
    \sf{\mathcal{Z}}[\sf{x}[N-1]] - \sf{A}[N-2]\sf{\mathcal{Z}}[\sf{x}[N-2]] - \sf{B}[N-2]^-\sf{\mathcal{Z}}[\sf{u}[N-2]]- \sf{B}[N-1]^+\sf{\mathcal{Z}}[\sf{u}[N-1]]
    }}_{\sf{A}[s]}
    \sf{x}[s] =
    \underbrace{
     \iv{
    \sf{x}[init] \\
    \sf{s}[0] \\
    \vdots \\
    \sf{s}[N-1] 
    }}_{\sf{b}[s]}
\end{equation}
Similarly, the linear and second order conic constraints take the form:
\begin{equation}
\begin{gathered}
    \sf{h}[s] = \iv{\sf{h}[l,0]^\ts & \sf{h}[l,1]^\ts & \sf{h}[l,2]^\ts & \sf{h}[c,0]^\ts & \dots & \sf{h}[c,6]^\ts}^\ts
    ;\quad
    \sf{G}[s] = \iv{\sf{G}[l,0]^\ts & \sf{G}[l,1]^\ts & \sf{G}[l,2]^\ts & \sf{G}[c,0]^\ts & \dots & \sf{G}[c,6]^\ts}^\ts
    \\
    \sf{h}[l,i] = \iv{\sf{h}[l,i,0]^\ts & \dots & \sf{h}[l,i,k]^\ts & \dots & \sf{h}[l,i,N-1]^\ts}^\ts
    ;\quad
    \sf{G}[l,i] = \iv{\sf{h}[l,i,0]^\ts & \dots & \sf{G}[l,i,k]^\ts & \dots & \sf{h}[l,i,N-1]^\ts}^\ts
    ;\quad i\in\{0,1,2\};
    \\
    \sf{h}[c,j] = \iv{\sf{h}[c,j,0]^\ts & \dots & \sf{h}[c,j,k]^\ts & \dots & \sf{h}[c,j,N-1]^\ts}^\ts
    ;\quad
    \sf{G}[c,j] = \iv{\sf{h}[c,j,0]^\ts & \dots & \sf{G}[c,j,k]^\ts & \dots & \sf{h}[c,j,N-1]^\ts}^\ts
    ;\quad j\in\{0,\dots,6\};
\end{gathered}
\end{equation}
\begin{equation}
\begin{gathered}
    \underbrace{
     \iv{
    0 
    \\
    0
    }}_{\sf{h}[l,1,k]}
    -
    \underbrace{
        \iv{
        -\sf{\mathcal{Z}}[\sf{\gamma}[k]] \\
        -\sf{\mathcal{Z}}[\sf{\zeta}[k]] 
        }
    }_{\sf{G}[l,1,k]}
    \sf{x}[s] \in \mathbb{R}_+^2 
    ;\quad
    \underbrace{
     \iv{
    \ones{n_w}\\
    \ones{n_w} \\
    \ones{n_w} \\
    \ones{n_w} \\
    \ones{3}\\
    \ones{3} 
    }}_{\sf{h}[l,2,k]}
    -
    \underbrace{
    \iv{
    -\sf{\mathcal{Z}}[\sf{\tau}[k]]\\
    \sf{\mathcal{Z}}[\sf{\tau}[k]]\\
    -\sf{\mathcal{Z}}[\sf{h}[k]]\\
    \sf{\mathcal{Z}}[\sf{h}[k]]\\
    -\sf{\mathcal{Z}}[\sf{\omega}[k]]\\
    \sf{\mathcal{Z}}[\sf{\omega}[k]]
    }}_{\sf{G}[l,2,k]}
    \sf{x}[s] \in \mathbb{R}_+^{4 n_w + 6}
    ;\quad
    \underbrace{
     \iv{
    \delta_{xmax} \\
    \delta_{umax}
    }}_{\sf{h}[l,3,k]}
    -
    \underbrace{
    \iv{
    \sf{\mathcal{Z}}[\sf{\delta}[x,k]]\\
    \sf{\mathcal{Z}}[\sf{\delta}[u,k]]
    }}_{\sf{G}[l,3,k]}
    \sf{x}[s] \in \mathbb{R}_+^{2}
\end{gathered}
\label{eq:linear_ecos}
\end{equation}
\begin{equation}
\begin{gathered}
    \underbrace{
     \iv{
    \sqrt{1+\cos{\theta_{sun}}} \\
    \zeros[4][1]}}_{\sf{h}[c,1,k]}
    -
    \underbrace{
    \iv{
    0\\
    -\sf{M}(\sf{r}[sun](t_k),\sf{v})\sf{\mathcal{Z}}[\sf{q}[k]]
    }}_{\sf{G}[c,1,k]}
    \sf{x}[s] \in \mathbb{Q}^{5};\quad 
     \underbrace{
     \iv{
    0 \\
    \zeros[4][1]}}_{\sf{h}[c,2,k]}
    -
    \underbrace{
    \iv{
    -\sf{\mathcal{Z}}[\sf{\eta}[k]]\\
    -\sf{N}(\sf{r}[c](t_k),\sf{v})\sf{\mathcal{Z}}[\sf{q}[k]]
    }}_{\sf{G}[c,2,k]}
    \sf{x}[s] \in \mathbb{Q}^{5};\quad
    \\
    \underbrace{
     \iv{
    \sqrt{1+\cos{\theta_{vmax}}} \\
    \zeros[4][1]}}_{\sf{h}[c,3,k]}
    -
    \underbrace{
    \iv{
    -\sf{\mathcal{Z}}[\sf{\gamma}[k]]\\
    -\sf{N}(\sf{r}[c](t_k),\sf{v})\sf{\mathcal{Z}}[\sf{q}[k]]
    }}_{\sf{G}[c,3,k]}
    \sf{x}[s] \in \mathbb{Q}^{5};\quad
    \underbrace{
    \iv{
    0 \\
    \zeros[n_w][1]}}_{\sf{h}[c,4,k]}
    -
    \underbrace{
    \iv{
    -\sf{\mathcal{Z}}[\sf{\rho}[k]]\\
    -\sf{\mathcal{Z}}[\sf{u}[k]]
    }}_{\sf{G}[c,4,k]}
    \sf{x}[s] \in \mathbb{Q}^{n_w + 1};
    \\
    \underbrace{
     \iv{
    \sqrt{1+\cos{\theta_{imax}}} \\
    \zeros[4][1]}}_{\sf{h}[c,5,k]}
    -
    \underbrace{
    \iv{
    -\sf{\mathcal{Z}}[\sf{\zeta}[k]]\\
    -\sf{N}(\sf{r}[c](t_k),\sf{v})\sf{\mathcal{Z}}[\sf{q}[k]]
    }}_{\sf{G}[c,5,k]}
    \sf{x}[s] \in \mathbb{Q}^{5};\quad
    \underbrace{
     \iv{
    0 \\
    \sf{\overline{u}}[k]}}_{\sf{h}[c,6,k]}
    -
    \underbrace{
    \iv{
    -\sf{\mathcal{Z}}[\sf{\delta}[u,k]]\\
    \sf{\mathcal{Z}}[\sf{u}[k]]
    }}_{\sf{G}[c,6,k]}
    \sf{x}[s] \in \mathbb{Q}^{n_w + 1};
    \quad
    \underbrace{
     \iv{
    0 \\
    \sf{\overline{x}}[k]}}_{\sf{h}[c,7,k]}
    -
    \underbrace{
    \iv{
    -\sf{\mathcal{Z}}[\sf{\delta}[x,k]]\\
    \sf{\mathcal{Z}}[\sf{x}[k]]
    }}_{\sf{G}[c,7,k]}
    \sf{x}[s] \in \mathbb{Q}^{7 + n_w + 1}
    \hspace{-20pt}
\end{gathered}\raisetag{5\baselineskip}
\label{eq:conic_ecos}
\end{equation}
In this form, the only matrices that are not fixed are the ones that depend on the linearization variables namely $\sf{c}[s],\, \sf{A}[s],\, \sf{b}[s],\, \sf{h}[c,6],\, \sf{h}[c,7],\, \sf{G}[c,6],\, \sf{G}[c,7]$ or those that impose the maximum trust region sizes i.e. $\sf{h}[l,3]$.
\subsubsection{Sequential programming algorithm}
The overall steps of the proposed guidance method can now be outlined in \cref{alg:scvx}.
\begin{algorithm}[ht]
    \DontPrintSemicolon
    \KwIn{initial linearization trajectories $\sf{\overline{x}}[k], \sf{\overline{u}}[k], \sf{\overline{\gamma}}[k], \sf{\overline{\zeta}}[k]\,\,\forall k \in \{0,\dots,N-1\}$}
    \KwOut{final guidance trajectory $\sf{x}[k], \sf{u}[k]$}
    compute the static parts of the matrices used in the convex solver call using \cref{eq:linear_ecos,eq:conic_ecos}\\
    \For{$i\leftarrow 1$ \KwTo $n_{iter}$}{
    compute linearization variables $\sf{A}[k], \sf{B}[k]^-, \sf{B}[k]^+, \sf{s}[k]$ using $\sf{\overline{x}}[k], \sf{\overline{u}}[k]$ and \cref{eq:discretization}\label{alg:scvx:linearization}\\
    update the parameters that depend on the linearization trajectory used in the convex solver call\\
    \For{$j\leftarrow 1$ \KwTo $n_{sol}$}{
    update the parameters that depend on the trust region size used in the convex solver call\\
    solve optimization \cref{eq:convex_optimization} to obtain $\sf{x}[k], \sf{u}[k], \sf{\gamma}[k], \sf{\zeta}[k], \sf{\delta}[x,k], \sf{\delta}[u,k]$\label{alg:scvx:solve}\\
    compute the true state trajectory $\sf{x}[k]^\star$ by integrating the input $\sf{u}[k]$ using \cref{eq:foh,eq:dynamics}
    \label{alg:scvx:integration}
    \\
    calculate the solution quality metric $\epsilon_x$ using $\sf{x}[k], \sf{x}[k]^\star$ and \cref{eq:epsilon}\\
    \If{$\epsilon_x \le \epsilon_{max}$}{
    expand trust region limits $\delta_{xmax},\,\, \delta_{umax} \leftarrow \kappa^+ \delta_{xmax},\,\, \kappa^+ \delta_{umax}$\\
    accept new linearization trajectories
    $\sf{\overline{x}}[k], \sf{\overline{u}}[k], \sf{\overline{\gamma}}[k], \sf{\overline{\zeta}}[k] \leftarrow
    \sf{x}[k]^\star, \sf{u}[k], \sf{\gamma}[k], \sf{\zeta}[k]$\\
    \Break\\
    \lElse{
    contract trust region limits $\delta_{xmax},\,\, \delta_{umax} \leftarrow \kappa^- \delta_{xmax},\,\, \kappa^- \delta_{umax}$
    }
    \lIf{time limit exceeded \Or $j = n_{sol}$}{\Return $\sf{\overline{x}}[k], \sf{\overline{u}}[k]$}\label{alg:scvx:termination0}
    }
    }
    \If{$\sum\limits_{k=0}^{N-1}\left(\sf{\delta}[u,k] + \sf{\delta}[x,k]\right) \le \delta_{convergence}$\label{alg:scvx:convergence_test}}{\Return $\sf{\overline{x}}[k], \sf{\overline{u}}[k]$}
    \lIf{time limit exceeded \Or $i = n_{iter}$}{\Return $\sf{\overline{x}}[k], \sf{\overline{u}}[k]$}\label{alg:scvx:termination}
    }
    \caption{Comet Interceptor Flyby Guidance using Sequential Convex Programming}
    \label{alg:scvx}
\end{algorithm}
\Cref{alg:scvx:convergence_test} in the algorithm tests convergence by checking if the sum of the all state $\sf{\delta}[x,k]$ and control deviations $\sf{\delta}[u,k]$ drops below a certain target threshold $\delta_{convergence}$. Since the algorithm is meant for real-time implementation, the number of maximum iterations is limited to $n_{iter}$. Similarly the maximum number of times that a solution can be rejected is also limited by $n_{sol}$. However, since the maximum trust region sizes decrease exponentially via the term $\kappa^-$, only a few contractions are typically needed to achieve a good solution quality metric $\epsilon_x$. \Cref{alg:scvx:termination0,alg:scvx:termination} help ensure that the algorithm returns a valid but sub-optimal solution if a certain time limit is exceeded or the maximum number of iterations has been reached.

\begin{remark}
 No theoretical convergence guarantees for the proposed algorithm are presented in this paper. However, the method worked well in practice as showcased in \cref{sec:results}. The convergence properties of similar sequential algorithms are discussed in \cite{Mao2016,Morgan2016,Lu2013c,Virgili-Llop2019a}. The method is particularly similar to the work from \cite{Mao2016} that relied on virtual control terms and hard trust regions to guarantee global convergence to a stationary but not necessarily feasible point. In this prior work, virtual control terms were used to alleviate the so-called artificial infeasibility that could result from the linearization of the dynamics or other highly nonconvex constraints. In this case, the linearized formulation could generate an infeasible problem, even if the original nonlinear problem is feasible.
 In this paper, virtual control were not used since the algorithm performed well in practice and artificial infeasibility was not deemed a major concern. Adding these extra terms would also significantly increase the number of decision variables and make on-board implementation on limited hardware even more challenging. The fact the dominant pointing constraints are all soft and the attitude dynamic and kinematic constraints \cref{eq:dynamics} are bilinear could be an explanation for why these virtual control terms were not needed in practice for this specific study case. 
\end{remark}
\begin{remark}
Neither the original optimization problem \cref{eq:nonlinear_optimization} nor its approximation \cref{eq:convex_optimization} directly enforce the unit norm non-convex constraint on the quaternion, i.e. $\vf{q}^\ts\vf{q} = 1$. However, the quaternion norm is indirectly constrained through the kinematic equation \cref{eq:dynamics}. Furthermore, every linearization in \cref{alg:scvx} is performed around the true state trajectory $\sf{x}[k]^\star$ obtained by integrating the nonlinear dynamics using the control inputs $\sf{u}[k]$ returned by the convex solver. Therefore, the errors introduced by neglecting the unit norm constraints on the quaternion remain very limited and these constraints can be safely ignored for performance reasons.
\end{remark}
\subsubsection{Initialization}
A good initial trajectory that satisfies all the constraints and is also close to the global optimum can speed up the convergence of \cref{alg:scvx}. However, generating such an initial trajectory was beyond the scope of this paper and instead the simple null torque input initial trajectory $\sf{\overline{u}}[k] = \zeros$ was used. The variables linearizing the cardinality objectives were set to unit values $\sf{\overline{\gamma}}[k] = \sf{\overline{\zeta}}[k] = 1$ for the first iteration.
As shown in \cref{sec:results}, the algorithm was able to successfully converge to good trajectories even with this simple initialization scheme.

\section{Numerical results}
\label{sec:results}
In this section we present a simulation case study to demonstrate the viability of the proposed approach for the Comet Interceptor flyby problem. Numerical values for the parameters of sequential convex algorithm and those describing the mission scenario and the spacecraft are provided in \cref{tab:parameter_values}. The code necessary to reproduce all of the results presented in this paper will be made accessible at \url{https://github.com/valentinpreda/scvx_comet_interceptor} upon publication of this work. The open source ECOS solver \cite{domahidi2013ecos} was used to carry out the conic optimization and the sparse matrices used for the call were assembled manually. A Python layer was used to connect the computationally intensive routines. Numerical integration of was performed using \texttt{LSODA} from the FORTRAN library \texttt{odepack} (\texttt{odeint} from \texttt{scipy.integrate} Python package). The relative and absolute error tolerances ($a_{tol}$ and $r_{tol}$) used for both the linearization (\cref{alg:scvx:linearization} in \cref{alg:scvx}) and final integration (\cref{alg:scvx:integration} in \cref{alg:scvx}) are provided in \cref{tab:parameter_values}. LU decomposition of the state transition matrix $\sf{\Phi}$ from \cref{eq:discretization} was performed using \texttt{GETRF} from \texttt{LAPACK} (\texttt{lu\_factor} from \texttt{scipy.linalg} Python package).

\subsection{Benchmark problem description}
\begin{figure}[ht]
    \centering
    \def\svgwidth{0.35\textwidth}
    {
    \footnotesize
    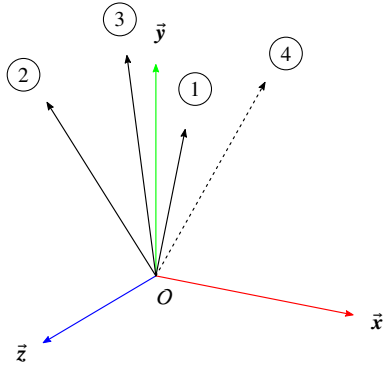
    }
    \caption{Spin-axis direction for each of the four wheels in the assembly. Dashed line indicates the faulty wheel.}
    \label{fig:wheel_arrangement}
\end{figure}

\begin{figure}[ht]
    \centering
    \includegraphics[width=0.35\textwidth, trim={5pt 5pt 5pt 5pt},clip]{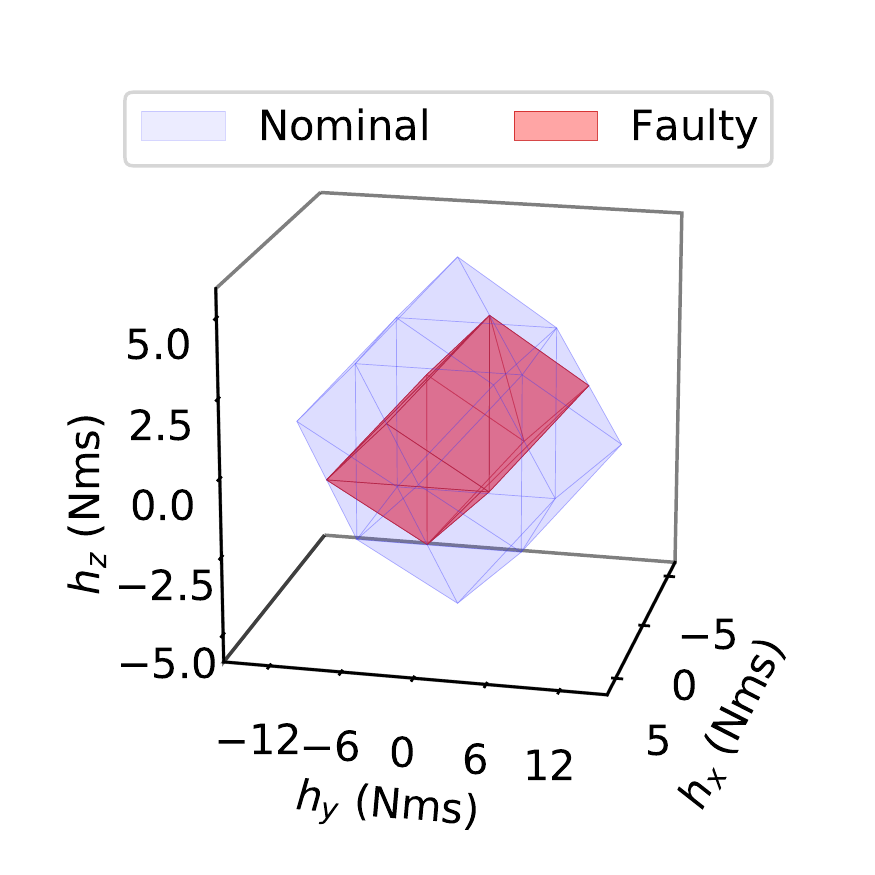}
    \caption{Set of achievable wheel momenta in nominal and faulty scenarios.}
    \label{fig:polytope_comparison}
\end{figure}

The benchmark problem consists of a linear flyby next to the comet target optimized over a  $\SI{200}{s}$ time interval such that the closest approach happens midway at $t = \SI{100}{s}$ at a distance of $\SI{1000}{km}$. The origin $O_\mathcal{I}$ of the inertial coordinate frame $\mathcal{F}_\mathcal{I} := \coordframe{O_\mathcal{I}}{\sf{\vec{x}}[\mathcal{I}]}{\sf{\vec{y}}[\mathcal{I}]}{\sf{\vec{z}}[\mathcal{I}]}$ is fixed to the comet position with the unit vector $\sf{\vec{x}}[\mathcal{I}]$ pointing along the flyby direction and $\sf{\vec{z}}[\mathcal{I}]$ towards the spacecraft at closest approach. At $t=0$, the spacecraft is assumed to have zero angular rate and its body axis vector $\sf{\vec{y}}$ is set perpendicular to the flyby plane as shown in \cref{fig:frames}. At the same time, the camera vector $\sf{\vec{v}}$ is aligned with the  $\sf{\vec{x}}$ body axis and with the comet pointing vector $\sf{\vec{r}}[c]$. Four reaction wheels are arranged in a pyramid configuration as shown in \cref{fig:wheel_arrangement}, with spin-axes tilted along the $\sf{\vec{y}}$ body axis to provide more capability about this primary slew axis. The maximum momentum and torque limits are set to match those of commercially available wheels and include an extra 20\% margin for additional attitude control system  authority.
Two scenarios are considered: a nominal case and a faulty case where the fourth wheel is assumed to be blocked, i.e. the fourth column in the torque distribution matrix $\sf{L}$ is eliminated and the maximum torque and momentum constraints are only applied for the remaining three active wheels. The direction of this fourth wheel is highlighted in \cref{fig:wheel_arrangement}, while \cref{fig:polytope_comparison} compares the set of achievable body frame momenta in both nominal and faulty scenarios. 

The primary uncertainty for the flyby is the dust field as it is highly comet dependent. The numerical study aims to evaluate the ability of the algorithm to recover science time after a large dust particle collision. To simulate such an impact, the wheel momentum vector $\sf{h}$ is initialized to different values from the convex set $\mathcal{H}$ covering 90\% of the each of the wheel's momentum range, i.e
\begin{equation}
    \sf{h}(0) \in \mathcal{H}\quad\text{with}\quad\mathcal{H}:=\left\{\sf{h} \in \mathbb{R}^{n_w}\;\;|\;\;|\sf{h}| \preccurlyeq 0.9\, \sf{h}[max]\right\} \subset \mathcal{H}_{max}
\end{equation}
where $n_w=4$ in the nominal case, $n_w=3$ in the faulty case and $\mathcal{H}_{max}$ denotes the convex set of achievable wheel momenta. This approach captures the momentum transferred to the spacecraft by the dust particle and assumes it was fully transferred to the wheels just prior to the final approach.
\begin{figure}[ht]
    \centering
    \includegraphics[width=\textwidth]{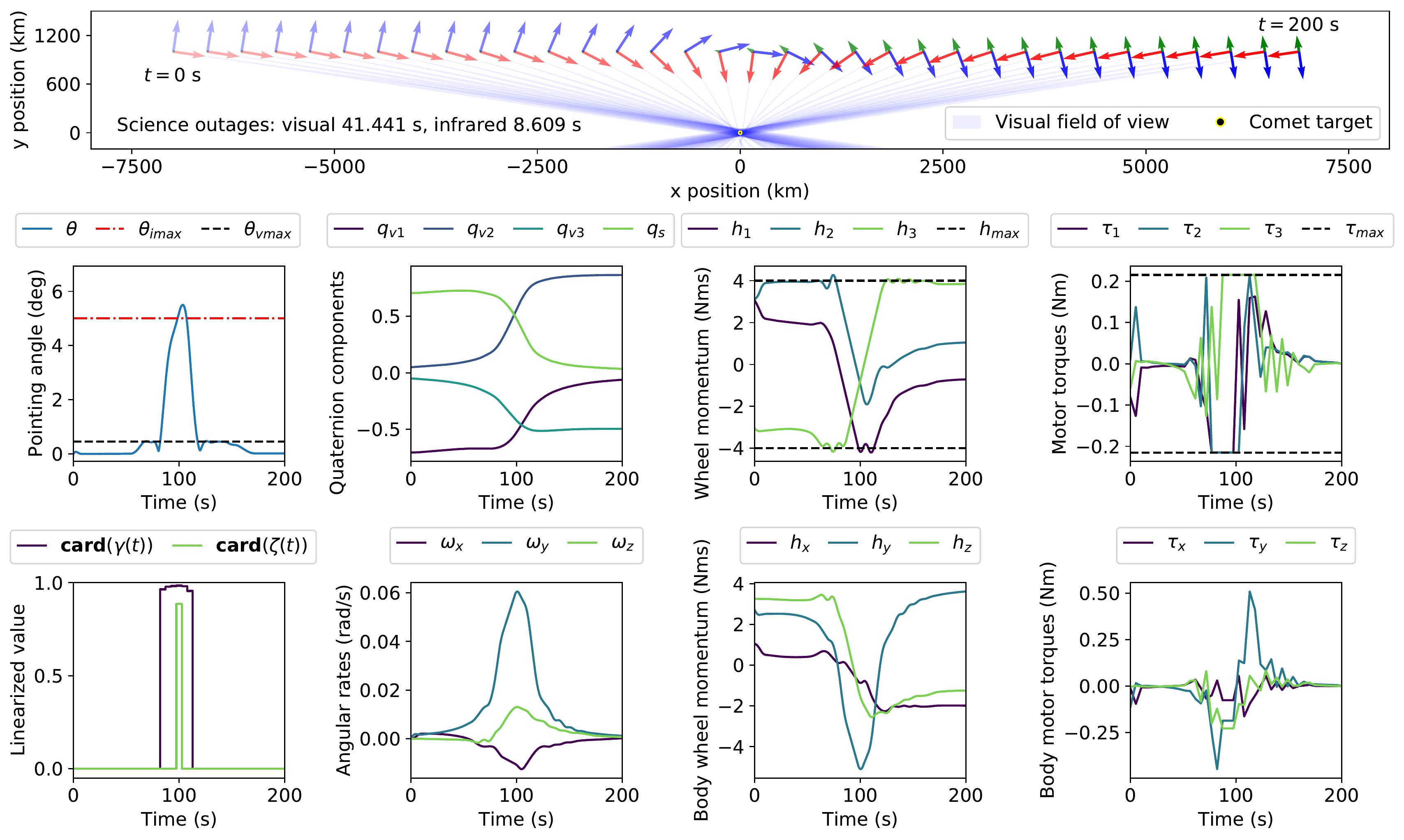}
    \caption{Flyby trajectory optimized for a faulty wheel scenario with the remaining wheels initially spinning close to saturation.}
    \label{fig:faulty_results}
\end{figure}

\begin{figure}[ht]
    \centering
    \includegraphics[width=0.3\textwidth, trim={5pt 5pt 2pt 5pt},clip]{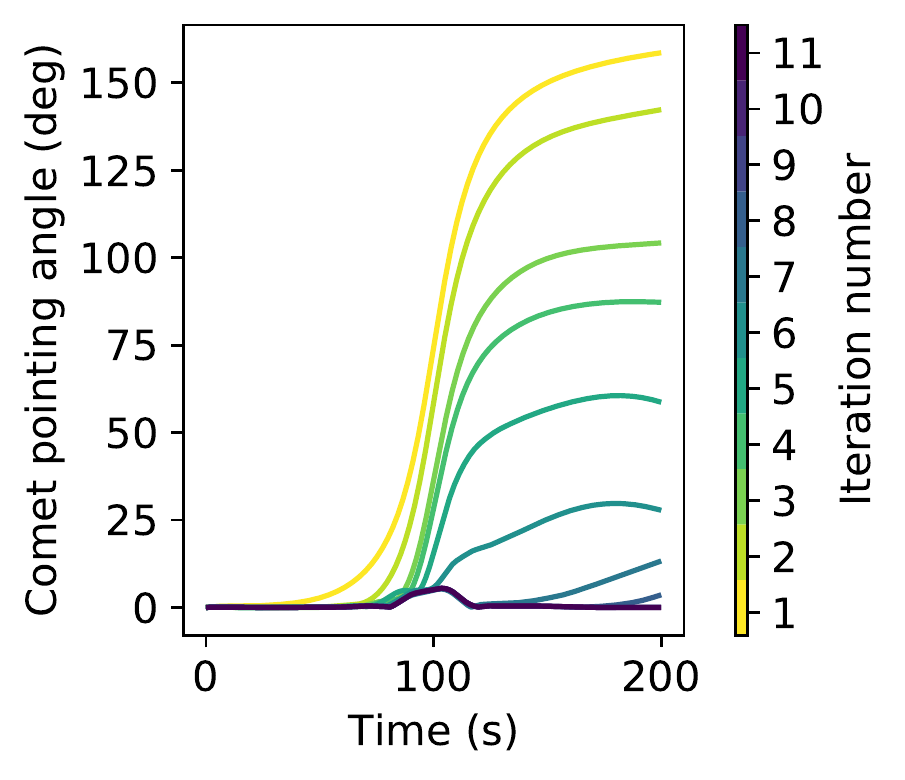}
    \caption{Pointing error angles at every iteration in the sequential optimization for the example shown in \cref{fig:faulty_results}.}
    \label{fig:angle_iterations}
\end{figure}


For clarity, \cref{fig:faulty_results}, shows an example trajectory returned by the proposed algorithm for a challenging faulty wheel scenario where all the remaining wheels are initially close to their maximum angular speed. The figure includes a number of important signal metrics such as the applied wheel torque in the body frame $\left[\tau_x(t)~ \tau_y(t) ~ \tau_z(t)\right]^\ts = \sf{L}\sf{\tau}(t)$, the wheel momentum in the body frame $\left[h_x(t)~ h_y(t) ~ h_z(t)\right]^\ts = \sf{L}\sf{h}(t)$ as well as the linearized and discretized values of the cardinality functions \cref{eq:card_approx} involved in the pointing objectives of the convex problem \cref{eq:convex_optimization}, i.e. $\sf{\gamma}[k]/(\varepsilon + \sf{\overline{\gamma}}[k])$ and $\sf{\zeta}[k]/(\varepsilon + \sf{\overline{\zeta}}[k])$. It is visible that these linearized cardinality values are close to unit value whenever the pointing angle exceeds either the visual or infrared field of view constraints and zero elsewhere. As expected these values are not always unit value because of the slight difference between $\sf{\gamma}[k], \sf{\zeta}[k]$ and their linearization counterparts $\sf{\overline{\gamma}}[k], \sf{\overline{\zeta}}[k]$. However, the strategy is effective at pushing the comet pointing angle below the field of view constraints. For this example trajectory, \cref{fig:angle_iterations} shows the different comet pointing angles obtained at every iteration in \cref{alg:scvx}. As the trajectory was initialized with a null control signal, the first iterations have a high angular error towards the end of the flyby. However, these errors are quickly penalized between each of the iterations leading to the final trajectory shown in \cref{fig:faulty_results}.
\subsection{Monte-Carlo campaign results}
To get a better understanding of the statistical behavior of the proposed approach, 50000 trajectories were generated for both nominal and faulty scenarios with random initial wheel momenta drawn uniformly from the set $\mathcal{H}$.

\begin{figure}[ht]
\centering
\includegraphics[width=0.57\textwidth]{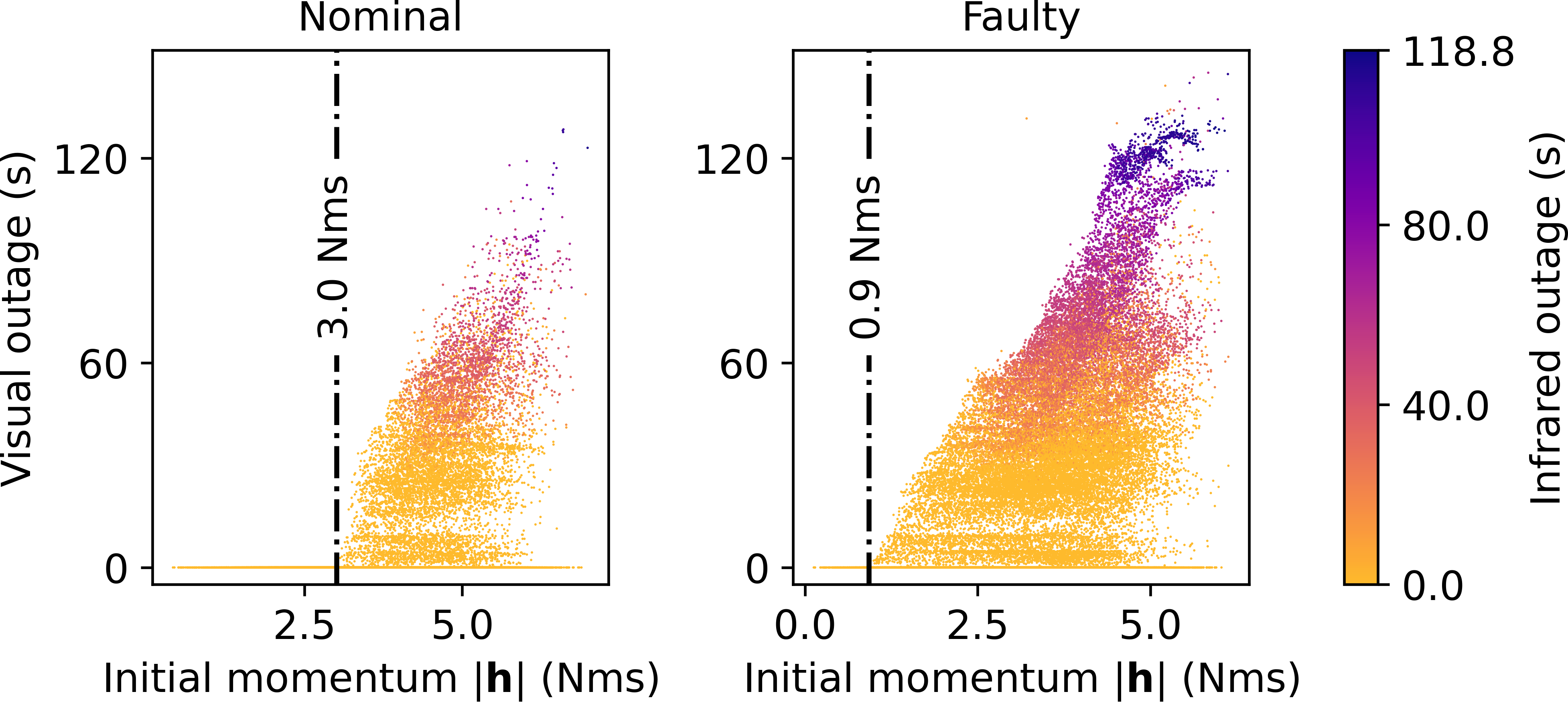}
\caption{Monte Carlo results: science outage times as a function of the initial wheel momentum in nominal and faulty conditions.}
\label{fig:momentum_vs_time}
\end{figure}

\begin{figure}[ht]
\centering
\includegraphics[width=0.37\textwidth]{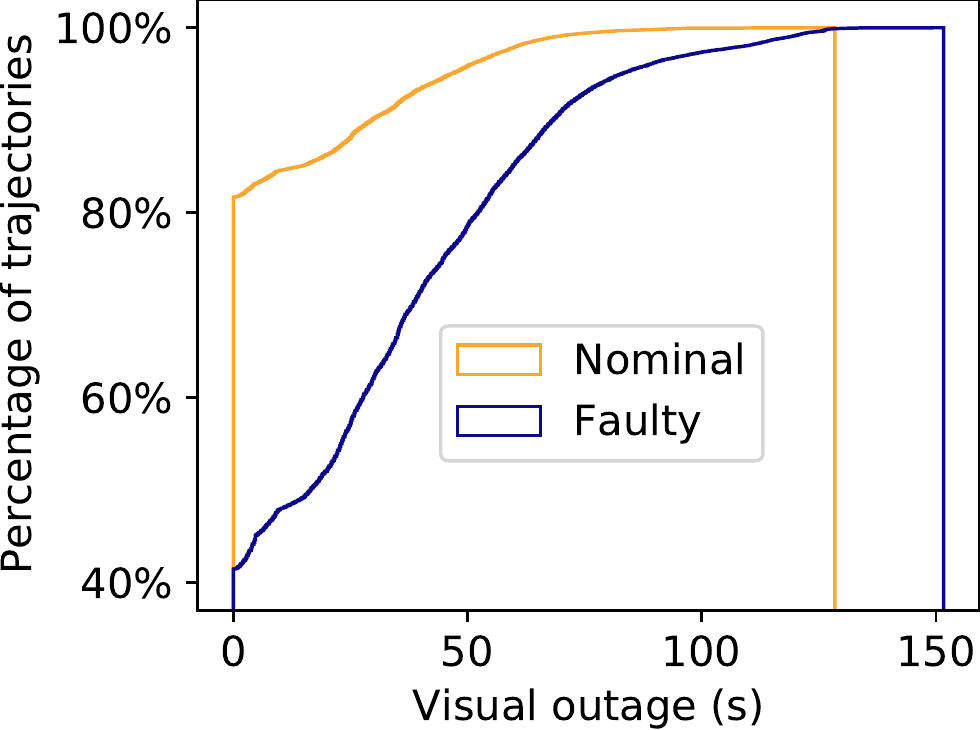}
\caption{Cumulative histogram of the visual outage results.}
\label{fig:visual_outage_rates}
\end{figure}

\begin{figure}[ht]
\centering
\includegraphics[width=0.38\textwidth]{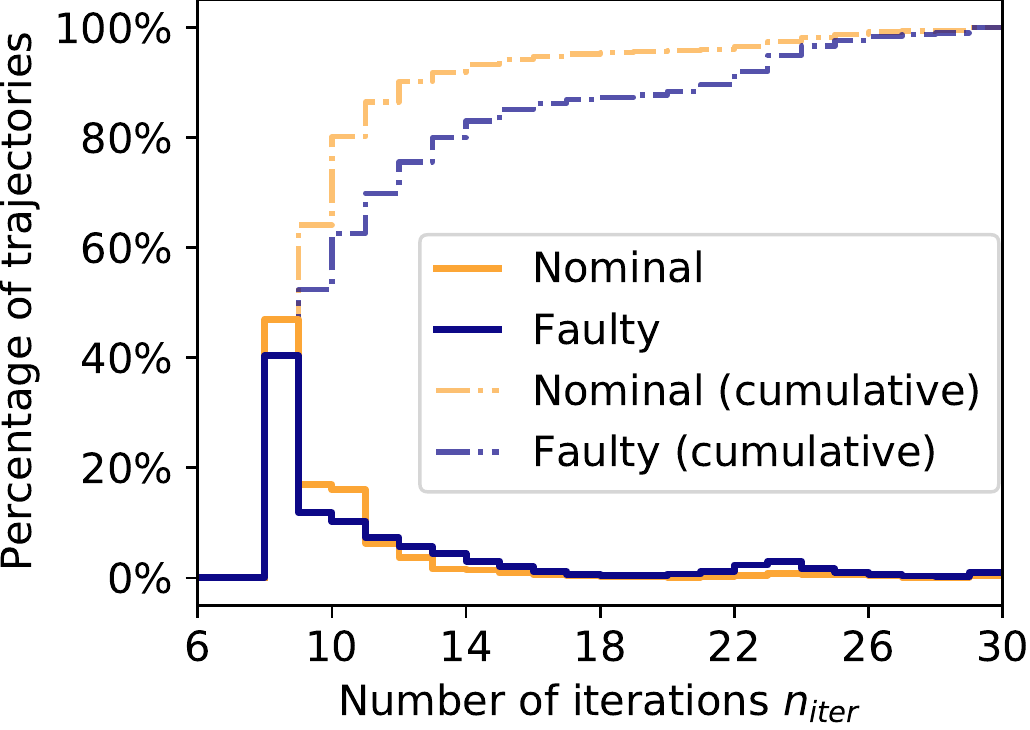}
\caption{Histograms and cumulative histograms of the number of iterations $n_{iter}$ in \cref{alg:scvx} needed for convergence.}
\label{fig:convergence_rates}
\end{figure}

\begin{figure}[ht!]
\centering
\includegraphics[width=\textwidth]{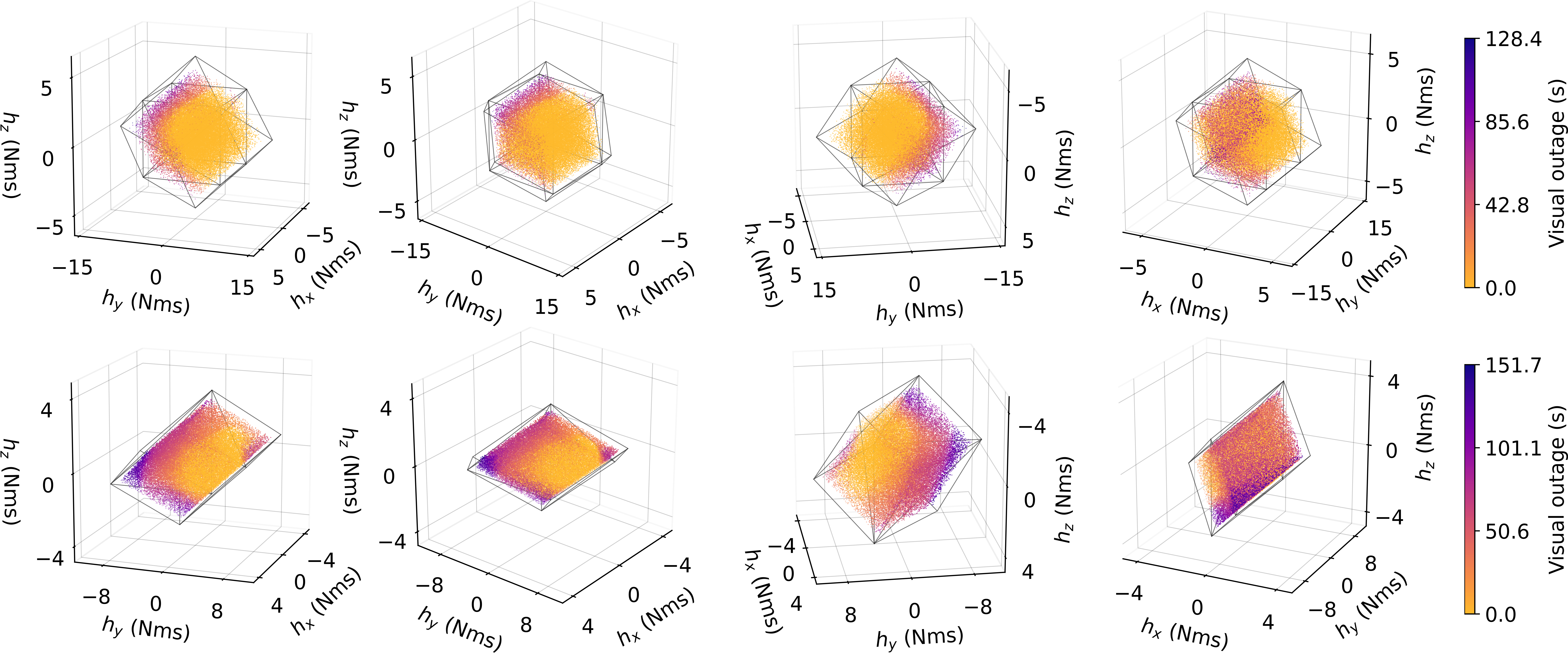}
\caption{Monte Carlo results: visual outage as a function of the initial body frame wheel momentum. Top row: nominal conditions; bottom row: faulty conditions.}
\label{fig:monte_carlo}
\end{figure}

\Cref{fig:momentum_vs_time} shows the distribution of the visual and infrared outage as a function of the initial wheel momentum magnitude. \Cref{fig:visual_outage_rates} displays a cumulative histogram of the visual outage pointing performance results, highlighting the percentage of trajectories that are below a certain threshold. As can be seen, in nominal conditions, more than 81.6\% of the initial conditions and all of those with $|\sf{h}(0)| \le \SI{3}{N.m.s}$ resulted in trajectories with no science outages. In faulty scenarios, these numbers degrade to $41.4\%$ and $\SI{0.9}{N.m.s}$.
The results reveal that the degradation in the worst-case performance follows almost linearly with the increase in the norm of the initial momentum. \Cref{fig:convergence_rates} includes a normal and a cumulative histogram of the number of iterations $n_{iter}$ needed for convergence. Less than 15 iterations were needed for more than $93\%$ of trajectories in the nominal case and $83\%$ in the faulty case. More than 25 iterations were needed for less than $0.8\%$ of nominal cases and less than $1.7\%$ of faulty cases. 
\Cref{fig:monte_carlo} further explores the performance degradation of the algorithm by plotting the visual outage as a function of the initial components of the body frame wheel momentum $\sf{L}\sf{h}(0)$. The figure also includes the convex hull defining the set of achievable wheel momenta in the body frame, i.e. $\mathcal{H}_{bmax} := \left\{\sf{L}\sf{h} \in \mathbb{R}^{3}\;\;|\;\;\sf{h} \in \mathcal{H}_{max}\right\}$. The results indicate that high visual outages occur mainly in along the boundaries of $\mathcal{H}_{bmax}$ with large negative values in $h_y$. This is consistent with physical intuition since for these initial conditions, the spacecraft doesn't have enough authority to slew around it's $\sf{\vec{y}}$ axis and keep the comet within the field of view at all times. Additionally, the pointing performance degrades smoothly as one approaches the boundaries. The results are encouraging since they indicate that the performance is mainly limited by physics and not by abnormal behavior in the algorithm. 

\Cref{tab:timing_values} summarizes the run time performance of different steps in \cref{alg:scvx} on multiple computing platforms computed across 1000 Monte Carlo runs. These preliminary results demonstrate that the algorithm is capable of executing even on very low-end hardware with decent performance. The run time of the linearization (\cref{alg:scvx:linearization} in \cref{alg:scvx}) and integration (\cref{alg:scvx:integration}) steps could be reduced in future iterations of the algorithm by switching to different integration routines, exploiting the sparsity structure of the state transition matrix \cref{eq:phi} when computing its inverse via LU decomposition in \cref{eq:discretization}, or switching to a compiled language instead of Python. The runtime performance of the optimization step (\cref{alg:scvx:solve}) could be improved by switching to a more advanced solver or removing some inactive constraints during the iterations. These extra optimizations were beyond the scope of the current study and will be the focus on future work.

\begin{table}
	\centering
	\caption{Run-time performance of different steps in \cref{alg:scvx} across multiple platforms}
	\renewcommand\arraystretch{1.2}
	\small
	\begin{tabular}{l  l l   l l   l l }
		\toprule
		 \multirow{2}{*}{Platform} & \multicolumn{2}{l}{Linearization (\cref{alg:scvx:linearization})} & \multicolumn{2}{l}{Optimization (\cref{alg:scvx:solve})} & \multicolumn{2}{c}{Integration (\cref{alg:scvx:integration})}
		\\
		 & Mean & Range & Mean & Range  & Mean & Range \\ 
		 \midrule
		 Intel i7-4910MQ @ \SI{2.9}{GHz} & \SI{63}{ms} & \SIrange{48.79}{195.8}{ms} & \SI{45.7}{ms} & \SIrange{29.5}{174.6}{ms} & \SI{12}{ms} & \SIrange{7.37}{30.2}{ms} \\
		 Raspberry Pi 3 Model B & \SI{0.733}{s} & \SIrange{0.593}{1.214}{s} & \SI{0.662}{s} & \SIrange{0.445}{2.852}{s} & \SI{140.13}{ms} & \SIrange{89.16}{168.77}{ms} \\
		 Raspberry Pi Zero W & \SI{5.17}{s} & \SIrange{4.2}{10.6}{s} & \SI{2.84}{s} & \SIrange{1.93}{11.8}{s} & \SI{1.14}{s} & \SIrange{0.716}{3.26}{s} \\
        \bottomrule
	\end{tabular}
	\label{tab:timing_values}
\end{table}

\section{Conclusions}
\label{sec:conclusions}
This work has outlined a methodology to generate spacecraft reorientation trajectories during a challenging flyby mission inspired by ESA's Comet Interceptor mission. Pointing outages outside the field of view of the imaging instruments were minimized by solving a novel convex-cardinality optimization problem using a sequential convex programming approach. The method relies on iteratively linearizing and discretizing the nonlinear dynamics constraints and cardinality objectives around a previous trajectory. The resulting set of convex problem are efficiently solved using second-order conic optimization software. Trust region constraints and objectives are introduced to restrict the optimization to a region where the linearization is valid. Detailed descriptions of all the steps in the algorithm were provided together with an extensive discussion on the results of the Monte Carlo simulation campaign. The results indicate that the performance of the proposed approach, gracefully degrade as one approach the fundamental physical limits of the system. Future work will focus on further optimizing the various steps of the algorithm to make it more suitable for real-time implementation on current flight hardware. An additional tracking controller will also be designed to keep the spacecraft on the guidance profile. Finally, future studies will investigate the alternative mission concept involving a scan mechanism.
\begin{table}[ht]
	\centering
	\caption{Numerical values of the benchmark parameters}
	\renewcommand\arraystretch{1.2}
	\small
	\begin{tabular}{l l}
		\toprule
		Parameter & Value
		\\
		\midrule
		\multicolumn{2}{c}{Spacecraft}
		\\
		\midrule
		Initial relative position in the inertial frame $\sf{r}[c](0)$ & $\iv{7000 & -1000 & 0}^\ts \si{km}$ 
		\\
		Velocity in the inertial frame & $\iv{0 & 70 & 0}^\ts  \si{km/s}$
		\\
		Initial quaternion $\sf{q}(0)$ (zero comet pointing angle) & $\iv{-0.7 & 0.05 & -0.05 & 0.7}^\ts$
		\\
		Initial angular velocity $\sf{\omega}(0)$ & $\iv{0 & 0 & 0}^\ts \si{rad/s}$ 
		\\
		Moment of inertia $\sf{J}$ & $\iv{225 & 10 & -10 \\ 10 & 128 & 10 \\ -10 & 10 & 223}$ \si{kg.m^2}
		\\
		Reaction wheel torque distribution matrix $\sf{L}$ & $\frac{\sqrt{2}}{4}\iv{1 & -1 & -1 & 1 \\ \sqrt{6} & \sqrt{6} & \sqrt{6} & \sqrt{6} \\ 1 & 1 & -1 & -1}$
		\\
		\midrule
		\multicolumn{2}{c}{Constraints}
		\\
		\midrule
		Pointing angles $\iv{\theta_{vmax} & \theta_{imax} & \theta_{sun}}$ & $\iv{0.46 & 5 & 60}\si{deg}$ \\
		Maximum wheel motor torque $\sf{\tau}[max]$ & $0.172 \iv{1 & 1 & 1 & 1}^\ts\si{N.m}$ \\
		Maximum wheel angular momentum $\sf{h}[max]$ & $3.2\iv{1 & 1 & 1 & 1}^\ts\si{N.m.s}$ \\
		Maximum angular rates $\sf{\omega}[max]$ & $\iv{5 & 5 & 5}^\ts$ \\
		\midrule
		\multicolumn{2}{c}{Sequential convex optimization}
		\\
		\midrule
		Number of sample points $N$ & 40 \\
		Total time $t_f$ & \SI{200}{s} \\
		Initial trust region sizes $\iv{\delta_{xmax} & \delta_{umax}}$ & $\iv{0.1 & 0.1}$
		\\
		Trust region update parameters $\iv{\kappa^+ & \kappa^-}$ & $\iv{2 & 0.25}$
		\\
		Maximum number of iterations and recomputations $\iv{n_{iter} & n_{sol}}$ & $\iv{30 & 20}$ \\
		Convergence threshold $\delta_{convergence}$ &  $10^{-2}$ \\
		Solution feasibility threshold $\epsilon_{max}$ &  $0.5$ \\
		 Integration tolerances $\iv{r_{tol} & a_{tol}}$ used for linearization & $\iv{10^{-5} & 10^{-5}}$ \\
		 Integration tolerances $\iv{r_{tol} & a_{tol}}$ computing the true state trajectory & $\iv{10^{-10} & 10^{-10}}$ \\
 		 Linearization term in the cardinality objectives $\varepsilon$ & $10^{-3}$ \\
		 \midrule
		 \multicolumn{2}{c}{Monte Carlo analysis} \\
		 \midrule
		 Number of random trajectories in both nominal and faulty scenarios & 50000 \\
		 Maximum absolute values of the initial momentum components $|\sf{h}(0)|$ & $0.9\,\sf{h}[max]$ 
		\\
		Index of the faulty reaction wheel & 4
		\\
        \bottomrule
	\end{tabular}
	\label{tab:parameter_values}
\end{table}
\bibliography{references}
\end{document}